\documentclass[pra,twocolumn,showpacs]{revtex4-1}

\usepackage{graphicx}
\usepackage{amsmath}
 \usepackage{amssymb}
\usepackage{amsfonts}
\usepackage{comment}
\usepackage{color}
\usepackage{hyperref}
\begin{document}

\title{Critical relaxation with overdamped quasiparticles in open
  quantum systems}

\author{Johannes Lang$^{1}$}
 \author{Francesco Piazza$^{2}$}

\affiliation{$^{1}$\small Physik Department, Technische Universit\"at M\"unchen, 85747 Garching, Germany}
\affiliation{$^{2}$\small Institut f\"ur Theoretische Physik, Universit{\"a}t Innsbruck, A-6020~Innsbruck, Austria}

\begin{abstract}
We study the late-time relaxation following a quench in a
open quantum many-body 
system. We consider the open Dicke model, describing the
infinite-range interactions between $N$ atoms and a single, lossy electromagnetic mode.
We show that the dynamical phase transition at a critical atom-light coupling is characterized by the interplay between reservoir-driven
and intrinsic relaxation processes in absence of number conservation. Above the critical coupling, small
fluctuations in the occupation of the dominant quasiparticle-mode
start to grow in time while the quasiparticle lifetime remains finite
due to losses.
Near the critical interaction strength we observe a crossover between exponential and power-law $1/\tau$
relaxation, the latter driven by collisions between
quasiparticles. For a quench exactly to the critical coupling, 
the power-law relaxation extends to infinite times, but the
finite lifetime of quasiparticles prevents ageing to appear in
two-times response and correlation functions. We predict our results to be accessible to quench experiments with ultracold bosons in optical resonators.
\end{abstract}

\maketitle

In a closed system, the relaxation toward the equilibrium state is
governed by processes which break integrability, allowing for an
efficient redistribution of energy and momentum between the degrees of
freedom. In this respect, important differences arise between classical and
quantum systems \cite{polkov_rev_2011,gogolin_2015_nature}. 
By contrast, in an open system the relaxation toward equilibrium
is driven by exchange of energy and momentum with an external
reservoir, so that the integrability-breaking intrinsic to the system
does not necessarily play a role in the late-time dynamics close to the
stationary state. In driven, dissipative systems, the latter is also generically different from a
thermal-equilibrium state, since detailed balance is usually violated. Moreover,
the presence of quantum correlations allows for the existence of
entangled stationary pure states determined by the reservoir \cite{diehl2008quantum}.
The scenario becomes even richer if one considers the relaxation
dynamics close to a phase transition. Already
for classical systems the standard theory of critical dynamics near equilibrium phase transitions \cite{hohenberg1977theory} does not fully characterize the relaxation
after quenches, since ageing-like behavior violates detailed-balance
\cite{calabrese_ageing_2006}. The extension of these concepts to quantum
\emph{and} open systems constitutes a challenging task which has recently received
much attention both for the near-steady-state
\cite{millis_2006,fazio_2008,diehl_2010,dallatorre_2010,weitz_2010,kessler_2012,kirton2013nonequilibrium,sieberer_2013,tauber_2014,bonnes_2014,lesanovski_rydberg_2014,pohl_ryd_2014,kehrein_2015,ott_2015,maghrebi_2015,marino_2016}
and quench \cite{marino_noisy_2012,sirker_2013,barthel_2013,giedke_2013,rey_2013,schmalian_2014,piazza_QKE,schuetz_2014,schuetz_2015,kollath_2015,buchhold_lutt_2015}
dynamics, also due to remarkable experimental advances in the control
of hybrid systems involving phonons/photons coupled to ions
\cite{blatt2012quantum,britton2012engineered}, excitons \cite{carusotto_rev_2013}, superconducting circuits
\cite{hartmann2008laser,houck2012chip,schmidt2013circuit}, mechanical
modes \cite{ludwig2013quantum}, or neutral atoms \cite{rauschenb_2010,cavity_rmp,kimble_2014_crystal,chang_many_body_2015}.

In this work, we consider an open quantum many-body system close to a
phase transition, where the interplay between dissipation and
integrability breaking in absence of number conservation gives rise to a novel scenario for the post-quench relaxation dynamics.
We consider an open version
\cite{carmichael_2007,simons_2010,bhaseen_2012,tureci_2012,domokos_open_fs_2012,domokos_open_ce_2011,dallatorre_2013,gritsev_driven_2013,lesanovski_dicke_ions,domokos_keldysh_2015,acevedo_2015}
of the paradigmatic Dicke model \cite{dicke_54}, describing
$N$ two-level atoms equally coupled to a single, lossy mode of the
electromagnetic field
\cite{lieb_1973,hioe_1973,emary_2004,vidal_2006,liu_finite_2009,larson_lew_2009,nagy_2010,larson_imp_2014},
recently realized experimentally with atoms in optical cavities
\cite{eth_2010,eth_2011,eth_2013,cavity_rmp,barrett_2014,hemmerich_2014,hemmerich_2015}. 
Due to the infinite range of the atom-photon interactions ($0$-dimensionality), this model is
integrable in the thermodynamic limit: $N=\infty$, corresponding to
non-interacting polaritonic quasiparticles. 
Despite the absence of local degrees of freedom (typically used to
characterize equilibration \cite{essler_2013}), integrability breaking
in the Dicke model at finite $N$ has been shown to lead to chaotic behavior \cite{emary_2003,fehske_2013}
and thermalization \cite{haake_2012} in the closed-system case. Thermalization can also be achieved at $N=\infty$ via disorder \cite{buchhold_dis_2013}.
Here we describe the late-time dynamics following a quench of the
atom-light coupling strength in the open system at finite $N$.
We show that quantum non-equilibrium fluctuations induced by quasiparticle interactions
trigger a dynamical phase transition
, which causes the occupation of the dominant quasiparticle-mode to become unstable and grow in time. However, the quasiparticle lifetime remains finite in presence of the Markovian losses. 
In the critical regime, we predict a
crossover between exponential and power-law $1/\tau$ relaxation. The
latter is driven by quasiparticle collisions and extends to infinite
times for a quench exactly to the critical point. However, since
the quasiparticles involved retain a finite lifetime throughout the
transition, the equilibration time does not diverge, thus ageing is
not observed in two-times functions. 

The algebraic dynamics with
overdamped quasiparticles is a genuine out-of-equilibrium many-body effect,
\emph{not} related to critical slowing down since the system size $N$ is finite.
The description of the relaxation driven by quasiparticles collisions
requires non-perturbative many-body techniques. In particular, it
cannot be described using mean-field approaches.

Quench experiments performed recently in the open Dicke model \cite{hemmerich_2015} have started exploring the dynamical phase transition, for which our theory provides the quantum description of the critical relaxation. Our predictions will be observable in the late-time behavior of response and correlation functions after small quenches near the critical point.

\section{The model} 
The Dicke model \cite{dicke_54} describes the coupling of $N$ two-level atoms to a
single mode of the electromagnetic field with the Hamiltonian 
\begin{align}
\hat{H} = \omega_0 \hat{a}^\dagger \hat{a} + \omega_z \hat{S}_z +\frac{2 g}{\sqrt{N}} \hat{S}_x \left(\hat{a}^\dagger + \hat{a}\right).
\end{align}
Here $\hat{S}_{z,x}=\frac{1}{2}\sum_{i=1}^{N}\sigma_i^{z,x}$ are collective
spin operators with the single-atom Pauli matrices $\sigma_i^{z,x}$
and $\hat{a}^\dagger$ and $\hat{a}$ are the bosonic photon creation and annihilation operators. 
$\omega_0$ is the characteristic photon frequency, $\omega_z$ the
splitting of the atomic levels and $g$ is the photon-atom coupling
strength. 
We will consider an open version of
this model by introducing Markovian photon losses with a rate $\kappa$. The non-unitary time evolution is described by the master equation for the density matrix $\rho$,
\begin{align}
\partial_t \rho = -i\left[\hat{H},\rho\right]+\kappa \left(2 \hat{a}\rho \hat{a}^\dagger - \left\{\hat{a}^\dagger \hat{a}, \rho \right\}\right).
\end{align}
Since we will be interested in large atom numbers $N$, we perform
a Holstein-Primakoff transformation: $S_z = -N/2+\hat{b}^\dagger
\hat{b}$ and $S^+ = \hat{b}^\dagger \sqrt{N - \hat{n}}\approx
\sqrt{N}\hat{b}^\dagger \left(1-\hat{n}/(2N)\right)$, while
$S_x=\frac{1}{2}\left(S^+ + S^-\right)$ and $S^-={S^+}^\dagger$,
yielding the following Hamiltonian 
 \begin{align}
 \label{eq:dm_hp}
 \begin{split}
 \hat{H}& = \hat{H}_0+\hat{H}'\\
 \hat{H}_0&= \omega_0 \hat{a}^\dagger \hat{a} + \omega_z \hat{b}^\dagger\hat{b}+g \left(\hat{a} + \hat{a}^\dagger\right)\left(\hat{b} + \hat{b}^\dagger\right)\\
 \hat{H}'&=- \frac{g}{2 N} \left(\hat{a} + \hat{a}^\dagger\right)\left(\hat{b}^\dagger \hat{b}^\dagger \hat{b} + \hat{b}^\dagger \hat{b} \hat{b}\right)+\text{O}\left(\frac{1}{N^2}\right).
 \end{split}
 \end{align}
For $N=\infty$ the interaction Hamiltonian $\hat{H}'$ vanishes and the model is integrable i.e. describes non-interacting quasiparticles
corresponding to polaritonic collective modes mixing atomic and
photonic excitations. 
This quadratic model has a superradiant transition \cite{lieb_1973,hioe_1973} at a critical coupling strength \cite{carmichael_2007,domokos_open_fs_2012,domokos_open_ce_2011,tureci_2012} 
\begin{align}
\label{eq:gc}
g_{c,0}=\sqrt{(\omega_0^2+\kappa^2) \omega_z/(4\omega_0)},
\end{align}
where a finite average polarization $\langle\hat{b}\rangle\propto
\sqrt{N}$ and a finite coherent light component $\langle\hat{a}\rangle\propto
\sqrt{N}$ spontaneously break the $\mathbb{Z}_2$ symmetry. The
transition is caused by a soft mode (see also
Fig. \ref{fig:kappa_sketch}) with zero characteristic frequency
$\omega_{\rm qp}$, which switches from being
damped to growing in time, i.e. the damping rate $\kappa_{\rm qp}$
crosses zero at $g_{c,0}$.
The transition is purely dissipative, i.e. characterized by completely
overdamped quasiparticles $\kappa_{\rm qp}\geq 0$ and $\omega_{\rm
  qp}=0$. This is due to the presence of Markovian losses while the transition is driven by the
Hamiltonian sector \cite{dallatorre_2013}.

The Hamiltonian (\ref{eq:dm_hp}) does not conserve the
excitation-number since it contains counter-rotating terms. This
has the same effect as a driving term, which can indeed compensate the
effect of losses, resulting in a steady state with a finite
excitation number
\cite{carmichael_2007,bhaseen_2012,tureci_2012,dallatorre_2013}. Moreover,
as it is the case in driven-dissipative systems,
the coexistence of counter-rotating terms and Markov losses violates
the detailed balance characterizing global equilibrium (see
\cite{dallatorre_2013} and Section \ref{sec:results}).

We conclude this section by pointing out that the absence of a
continuum (or extensive number) of degrees of freedom does not prevent
the system to show many-body behavior.
The Dicke model, due to the infinite range of atom-light interactions,
is 0-dimensional i.e. the spatial structure is lost. It therefore
describes many quasiparticle excitations occupying the 4 possible polaritonic
collective modes. The non-integrable model $N<\infty$ includes
interactions between these quasiparticles. Given the unlimited
Hilbert space in every mode and since the occupation numbers are
generically large ($O(N^{1/2})$) in the scaling regime, see
\cite{dallatorre_2013} and Section \ref{sec:results}), there is no
notion by which the system describes a few-body or impurity
problem. In particular, for the critical late-time dynamics of the system the
relaxation i.e. redistribution of energy between the modes is strongly
affected by quasiparticle collisions. This behavior cannot be
described using mean-field approaches and rather requires many-body
techniques as the non-perturbative diagrammatics introduced next.

\section{Approach}
The non-equilibrium critical properties of the open Dicke model have
been recently investigated in near-steady-state
\cite{eth_2013,tureci_nearss,domokos_damping} and quench
\cite{hemmerich_2015,bhaseen_2012} experiments. Here we want to go
beyond the semiclassical studies and describe the critical post-quench
late-time relaxation including quantum fluctuations due to
quasiparticle interactions at finite system sizes as well as classical fluctuations from the
Markov reservoir. We adopt a diagrammatic technique based on the real-time Keldysh functional-integral formulation of the
Dyson equation \cite{kamenev_book,sieberer2015keldysh}, extending the steady state approach
developed in \cite{dallatorre_2013} to include the relaxation induced by
quasiparticle collisions as well as the breaking of time-translation
invariance. 
In the Keldysh functional-integral approach \footnote{See
  Supplemental Material}, one derives the two coupled Dyson equations for the retarded
and Keldysh Green's function (GF):
\begin{align}
&G^K=G^R \circ\left(\Sigma^K-D_0^K\right) \circ G^{R^\dag} \label{eq:dyson_K}\\
&\left(\left[G^R_0\right]^{-1}-\Sigma^R\right)\circ G^R=\delta(t-t'),\label{eq:dyson_R}
\end{align}
where \textquotedblleft $\circ$\textquotedblright\;indicates
the convolution in real time. Due to the absence of number
conservation in the Hamiltonian, the GFs are 4 by 4 matrices:
$
i\left(G^K(t,t')\right)_{\rm i,j}=\langle \{ \hat{V}_{\rm i}(t),\hat{V}_{\rm j}^\dag(t')\}\rangle
$
and
$
i\left(G^R(t,t')\right)_{\rm i,j}=\theta(t-t')\langle [ \hat{V}_{\rm i} (t),\hat{V}_{\rm j}^\dag(t')]\rangle
$,
with $\hat{V}^T=(\hat{a},\hat{a}^\dag,\hat{b},\hat{b}^\dag)$.

The retarded GF encodes the spectral response of the
system, the Keldysh GF its
correlation functions. As detailed-balance cannot be assumed, we
must determine $G^R$ and $G^K$ independently through
Eqs.~\eqref{eq:dyson_K},\eqref{eq:dyson_R}. 
The retarded GF $G_0^R$ and the matrix $D_0^K$
are fixed by the non-interacting theory $\hat{H}'=0$ and 
given in the Appendix \ref{app:keldysh}.
Finally, the self-energies
$\Sigma^{(K,R)}=\Sigma^{(K,R)}\!\!\left[G^R,G^K\right]$ are computed
within a self-consistent Hartree-Fock (SCHF) approximation
(as for instance employed to describe spin-chain dynamics \cite{marino_2013}), 
corresponding to the selection of Feynman diagrams shown in Fig.~\ref{fig:diagrams}. Self-consistency is necessary to treat
late-time relaxation close to the steady state \cite{kamenev_book}. This is true despite
the presence of the Markov reservoir since the system is close to a phase transition.
Moreover, the inclusion of the Fock processes we perform
here is required to describe the effect of quasiparticle collisions
on the late-time relaxation of the system after a quench.

\section{Results} 
\label{sec:results}
Starting from an initial atom-photon coupling $g_i$, we consider a sudden quench to a value $g>g_i$. We solve the
coupled Dyson Eqs.~\eqref{eq:dyson_K},\eqref{eq:dyson_R} in the SCHF
approximation in the limit of large absolute times $\tau=(t+t')/2$,
i.e. for small relative deviations from the steady state, by
means of an iterative numerical procedure. This approximate time-evolution is
illustrated in detail in the Appendix \ref{app:numerics}.
In the limit of relative times $t_{\rm rel}$ long compared to the quasiparticle lifetime $1/\kappa_\text{qp}$, that is, including
only the dominant contribution from
low-frequency quasiparticles, the solutions
take the following form
\begin{align}
\label{eq:GF_longtime}
&\left(G^K(t_\text{rel},\tau)\right) _{\rm i,j} \simeq e^{-\kappa_\text{qp}|t_\text{rel}|}\nonumber\\
&\times\left(G^K(0,\infty)+\frac{\delta G^K(0,0)
  }{e^{\kappa_\text{kin}\tau}+\frac{\lambda_\text{kin}}{\kappa_\text{kin}}\delta G^K(0,0)\left(e^{\kappa_\text{kin}\tau}-1\right)}\right),\nonumber\\
&\left(G^R(t_\text{rel},\tau)\right) _{\rm i,j} \simeq \theta\left(t_\text{rel}\right)  e^{-\kappa_\text{qp}t_\text{rel}}\nonumber\\
&\times\left(G^R(0,\infty)+\frac{\delta G^R(0,0)
 }{e^{\kappa_\text{kin}\tau}+\frac{\lambda_\text{kin}}{\kappa_\text{kin}}\delta G^K(0,0)\left(e^{\kappa_\text{kin}\tau}-1\right)}\right),
\end{align}
where we used the notation $\delta G^{R/K}(0,0)=G^{R/K}(0,0)-G^{R/K}(0,\infty)$.
\begin{figure}[htp]
\includegraphics[width=\columnwidth]{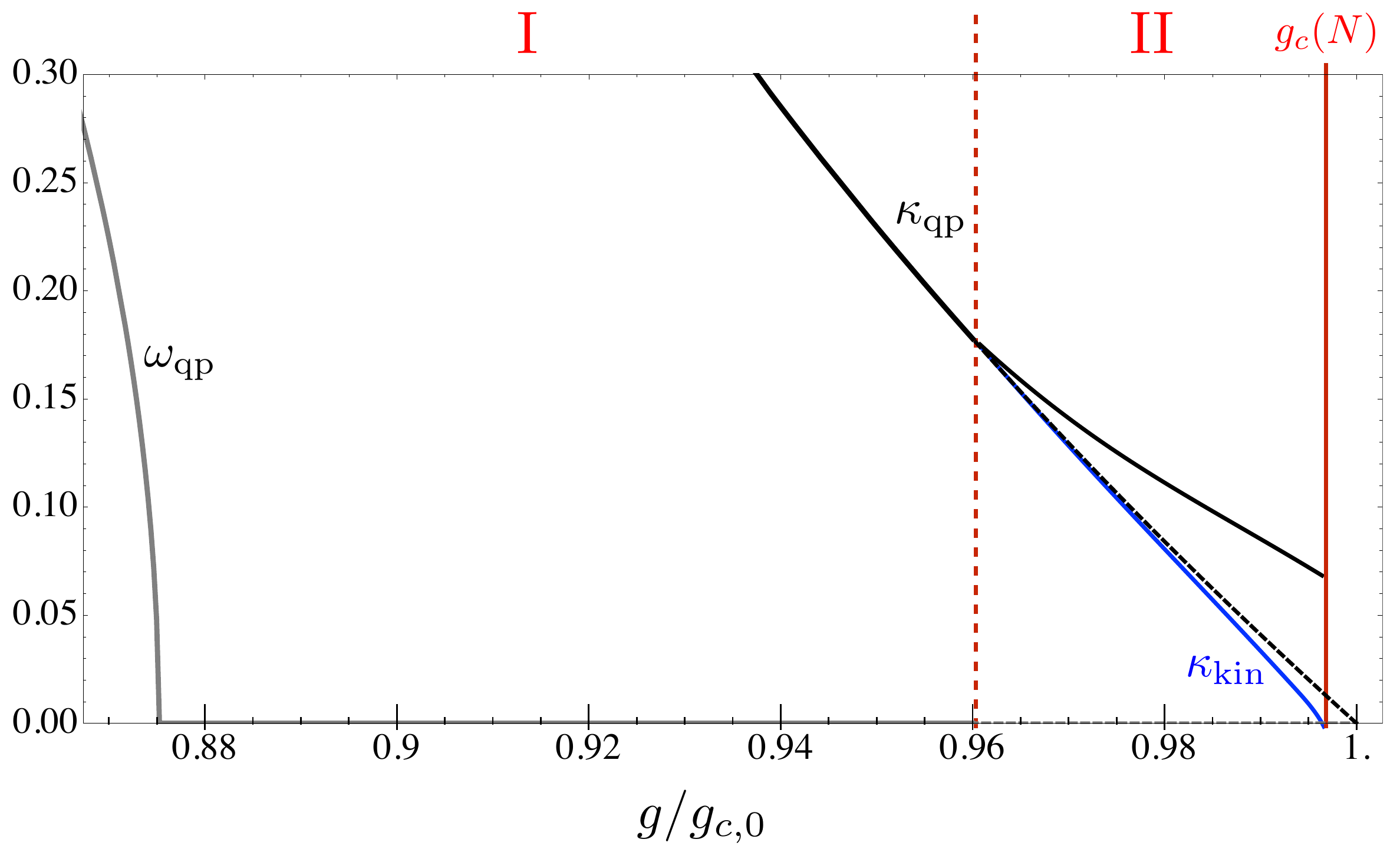}
\caption{Qualitative behavior of the quasiparticle characteristic
  frequency $\omega_{\rm qp}$ (gray) and inverse lifetime
  $\kappa_{\rm qp}$ (black),
  together with the system's damping rate $\kappa_{\rm kin}$ (blue), as a
  function of the final value $g$ of the light-matter coupling after
  a sudden quench from $g_i<g$. The dashed line corresponds to the
  prediction of the non-interacting theory $\hat{H}'=0$, where
  $\kappa_{\rm kin}=\kappa_{\rm qp}$. For $\omega_{\rm qp}$ there is
  no difference between interacting and non-interacting predictions at large enough $N$.
}
\label{fig:kappa_sketch}
\end{figure}
Every component ${\rm i,j}=1,...,4$ of both retarded and Keldysh GFs follows the
functional form (\ref{eq:GF_longtime}) since the latter is determined
by the least-damped quasiparticle mode corresponding to the dominant
eigenvector of the 4 by 4 matrices \footnote{Note that the single-mode approximation, implicit in the
  results presented here, breaks down for small coupling
  constants, where $\omega_\text{qp}\neq0$ and the two most relevant
  modes have degenerate lifetimes.}.
The solutions depend only on three parameters whose behavior is shown
in Fig.~\ref{fig:kappa_sketch} and \ref{fig:kappacrossing} as a
function of the coupling strength $g$: the quasiparticle
inverse lifetime $\kappa_{\rm qp}$ (damping the relative-time
dynamics), the system-damping $\kappa_{\rm kin}$ in the absolute time,
and the nonlinear coefficient $\lambda_{\rm kin}$.

\subsection{Dynamical phase transition at finite $N$}
\begin{figure}[htp]
\includegraphics[width=\columnwidth]{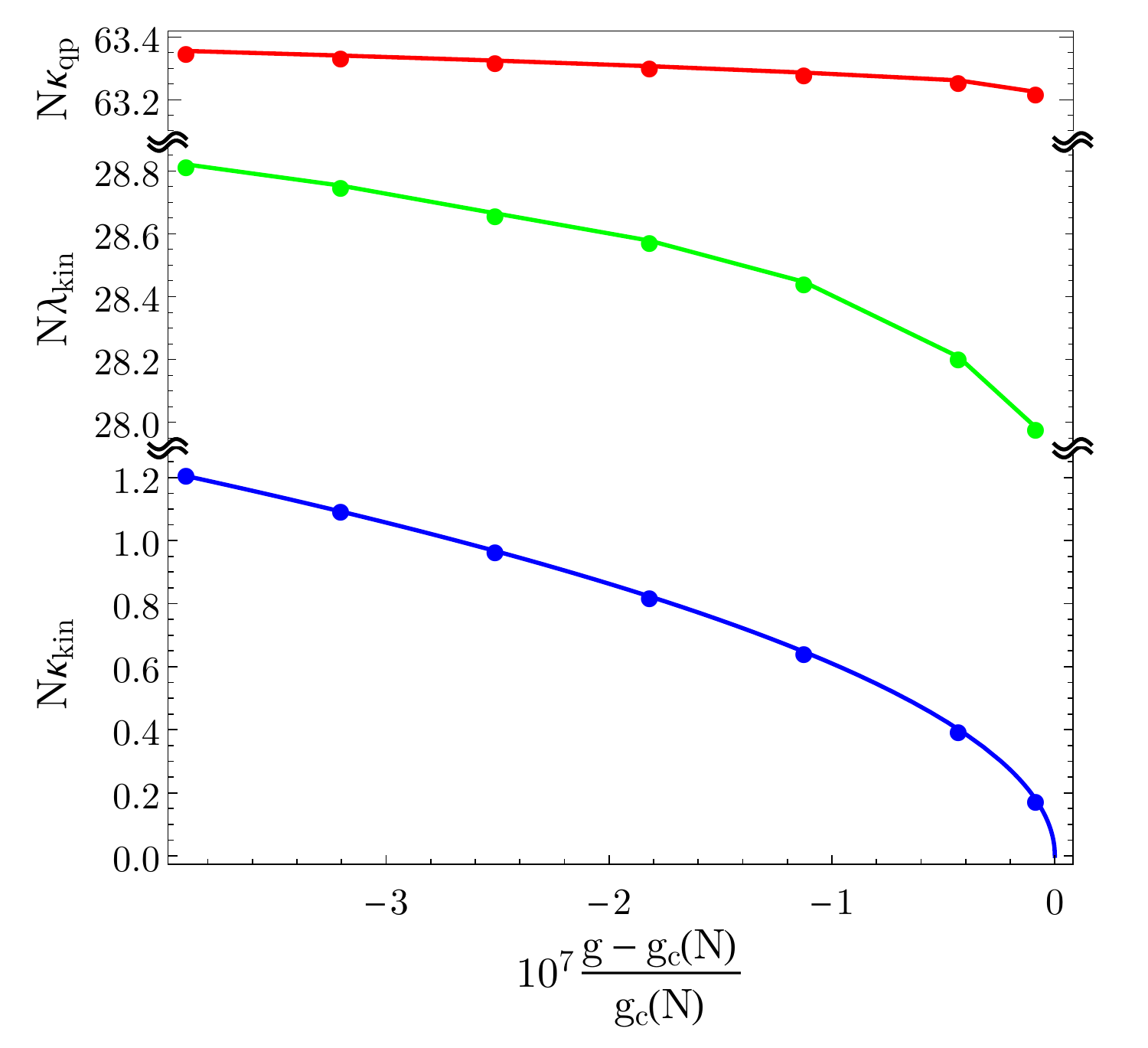}
\caption{Numerical values of the kinetic parameters $\kappa_{\rm kin}$,
  $\lambda_{\rm kin}$ together with the inverse
  quasiparticle lifetime $\kappa_{\rm qp}$. $\kappa_\text{kin}$ is fitted
  with the scaling law (\ref{eq:kkin}) (blue line). The parameters
  used are $\kappa=2$, $\omega_0=2$, $\omega_z=2.1$ and $N=1000$,
  resulting in $g_c\approx 1.4491$. In the Appendix \ref{app:kappa} we
  present results for $\kappa=0.2,1.0$.
}
\label{fig:kappacrossing}
\end{figure}

Let us first consider the integrable case:  $N=\infty$. Since the
quasiparticle interactions are absent, the system's damping is
equal to the quasiparticle damping: $\kappa_{\rm kin}=
\kappa_{\rm qp}$ and $\lambda_{\rm kin}=0$. Therefore $G^K(t_\text{rel},\tau) \simeq e^{-\kappa_\text{qp}|t_\text{rel}|}(G^K(0,\infty)+\delta G^K(0,0) e^{-\kappa_\text{qp}\tau})$ and analogously for the
retarded GF.
For $\tau\to\infty$ the steady state GF $G_{\rm ss}^K(t_\text{rel})
\simeq G^K(0,\infty) e^{-\kappa_\text{qp}|t_\text{rel}|}$ is
reached. 
As shown in Fig.~\ref{fig:kappa_sketch} by the black-dashed line, for
$N=\infty$ the inverse lifetime $\kappa_{\rm qp}$ vanishes linearly at the transition point
$g_{c,0}$. 
In the non-integrable $N<\infty$ case (solid lines in
Fig.~\ref{fig:kappa_sketch}), we find the phase transition to occur
instead at a critical coupling
\begin{align}
\label{eq:gcN}
g=g_c(N)<g_{c,0},\text{ with }\frac{|g_c(N)-g_{c,0}|}{g_{c,0}} \lesssim N^{-1/2},
\end{align}
where the inverse quasiparticle lifetime $\kappa_{\rm qp}$ remains
finite, while the damping
$\kappa_{\rm kin}$ vanishes according to:
\begin{align}
\label{eq:kkin}
\kappa_\text{kin}\sim \kappa_{\rm qp}N^{3/4}\sqrt{|g-g_c(N)|/g_c(N)}\;,
\end{align}
as shown in Fig.~\ref{fig:kappacrossing}. Above the critical point:
$g>g_c(N)$ the system's damping rate $\kappa_{\rm kin}$ becomes
imaginary, with the magnitude again given by \eqref{eq:kkin},
indicating an instability of the steady state of Eqs.~\eqref{eq:GF_longtime}.
This peculiar dynamical phase transition characterized by a vanishing system-damping at
finite quasiparticle lifetime is triggered by quasiparticle
collisions in presence of both Markovian losses and violation of
number conservation, the latter effectively working as a drive. 
In the following, we illustrate how this critical point affects the
system's dynamics after the quench.

\subsection{Criticality and scaling laws}
At any given $1\ll N<\infty$, sufficiently far away from the critical
point: $(g_c(N)-g)/g_c(N)\gtrsim N^{-1/2}$, we are in a weak-coupling regime (region I in Fig.~\ref{fig:kappa_sketch}) where the
quasiparticle interactions from $\hat{H}'$ are always perturbative
so that, to order $1/N$, the GFs follow the integrable dynamics illustrated above:
$\kappa_{\rm kin}\simeq\kappa_{\rm qp}\gg \lambda_{\rm kin}$.
Instead, for a quench to strong coupling $(g_c(N)-g)/g_c(N)\lesssim N^{-1/2}$
(region II in Fig.~\ref{fig:kappa_sketch}) the interactions
appreciably renormalize the dampings such that $\kappa_{\rm
  kin}<\kappa_{\rm qp}$. 
Within this region, even closer to the critical point: $(g_c(N)-g)/g_c(N)\lesssim
N^{-3/2}$, we find $\kappa_{\rm qp}\sim N^{-1/2}$ such that
$\lambda_{\rm kin}$ cannot be neglected any more (see
also Fig.~\ref{fig:kappacrossing}). In general, the latter depends only
weakly on the coupling $g$ and is also of order $N^{-1/2}$ \footnote{The
  scaling laws we found for $\kappa_{\rm kin},\kappa_{\rm qp},\lambda_{\rm
    kin}$, and $g_c(N)$ can be also derived from the scale-invariance of the
 GFs, which holds in the strong-coupling region. See J. Lang and F. Piazza, in preparation (2016)}. The role of $\lambda_{\rm kin}$ is to
introduce algebraic relaxation characteristic of
non-integrable dynamics. In our model without conserved quantities \cite{dallatorre_2013} algebraic dynamics emerges due to
criticality, but is in general not necessarily a signature of the
latter, for instance in systems with conservation laws \cite{hohenberg1977theory}.
At a given $N$-independent coupling $g$, the integrable
limit of the late time dynamics is reached for $N\to\infty$ since we
enter the weak coupling regime as soon as $(g_c(N)-g)/g_c(N)\gtrsim
N^{-1/2}$. If instead we pin the system to criticality $(g_c(N)-g)/g_c(N)\lesssim
N^{-3/2}$, the integrable limit is never approached since according to
\eqref{eq:kkin} $\kappa_{\rm qp}\simeq\kappa_{\rm
kin}\sim N^{-1/2}\to 0$ and $\lambda_{\rm kin}\sim N^{-1/2}\to
0$, so that the non-integrable character is always important. This
is related to the fact that at criticality the limits $N\to\infty$ and
$\tau\to\infty$ do not commute.
As a side remark, the fact that quasiparticle
collisions breaking integrability become important at criticality can be seen also by analyzing the steady
state. In particular, as shown in the Appendix \ref{app:F}, integrability breaking effectively creates a bath for the spin (atomic) degree of freedom. 

\subsection{Algebraic vs. Exponential dynamics}
\begin{figure}[t]
\includegraphics[width=\columnwidth,height=0.6\columnwidth]{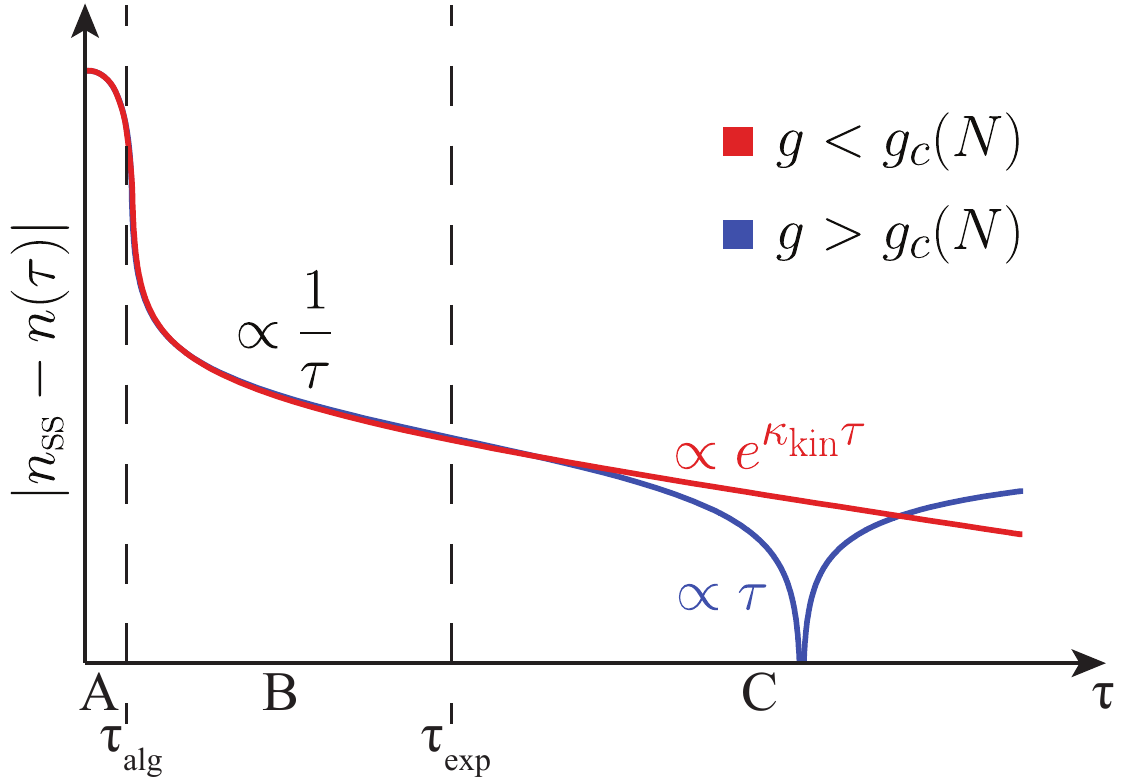}
\caption{Sketch of a log-plot of the time evolution of the
  particle number near $g_c(N)$, separated into three
  regions. Starting from the vacuum, the system is for short
  time-scales described by the evolution according to the bare Green's
  function (region A), which will then cross over into an algebraic
  decay (region B), that continues to infinite times for $g=g_c(N)$. For $g<g_c(N)$
  the final relaxation is exponential,
  as depicted in region C, whereas for $g>g_c(N)$ the population instead evolves linearly through that of the unstable steady state.}
\label{fig:late-time_dyn}
\end{figure}

An example depicting the generic behavior of the absolute-time evolution is sketched in
Fig.~\ref{fig:late-time_dyn} using the occupation of the quasiparticle
mode $n(\tau)= i G^K(0,\tau)/2-1/2$ as observable. After the quench the system
has to become sufficiently populated and correlated for interactions
to become important. This requires a time $\tau_{\rm alg}\sim
1/\kappa_{\rm qp}$, after which the
initial exponential integrable dynamics goes over into a
non-integrable  $1/\tau$ behavior. 
Deep inside the strong coupling regime: $|g_c(N)-g|/g_c(N)\lesssim
N^{-3/2}$, a second crossover takes place on a scale
$\tau_{\rm exp}\sim 1/\kappa_{\rm kin}$, where for $g<g_c(N)$ the
algebraic relaxation goes back to exponential, as predicted by
Eqs.~\eqref{eq:GF_longtime}. Using the result (\ref{eq:kkin}) we get the following scaling
\begin{align}
\tau_\text{exp}\sim 1/\kappa_\text{kin}\sim N^{-1/4}(|g-g_c(N)|/g_c(N))^{-1/2}\;.
\end{align}

The transcritical $g>g_c(N)$ time-evolution is also shown in
Fig.~\ref{fig:late-time_dyn}.
For times later than $\tau_\text{alg}$ the system first approaches the steady
state $G_{\rm ss}^{K,R}(t_\text{rel})$ of Eq.~\eqref{eq:GF_longtime} algebraically: $1/\tau$. However, beyond the
time-scale $\tau_\text{exp}$ the system then evolves linearly past this unstable state with a characteristic rate given by 
$\partial_\tau \delta O \simeq
\kappa^2_\text{kin}O_\text{ss}/\left(4\lambda_\text{kin}G^K_\text{ss}(0)\right)$
for any observable $O$. After the linear regime, the evolution
accelerates again, becomes algebraic and would eventually converge toward
the symmetry-broken steady state. The description of such a state however requires the expansion around a symmetry-broken saddle point, including (self-consistent) finite field expectation values $\langle \hat{a} \rangle$ and $\langle \hat{b}\rangle$, which is described by a more general version of Hamiltonian \eqref{eq:dm_hp}. The new steady state is therefore currently inaccessible to the presented dynamics.
The sudden switch in the dynamical behavior at $g=g_c(N)$
characterizing the phase transition is triggered by quasiparticle
collisions in presence of both Markovian losses and effective driving. In
particular, since the system has weakly damped
quasiparticles at $\omega_{\rm qp}=0$ (which is possible due to
Markovian losses), collisions take place almost on-shell and therefore
efficiently increase the mode occupation. The drive
(breaking number conservation) provides the source of quasiparticles
allowing the latter process to induce an instability.

\subsection{Absence of ageing}
\label{sec:ageing}
For a quench exactly to the critical point $g=g_c(N)$, the power-law
$1/\tau$ dynamics extends down to the steady state. 
Due to the breaking of time-translation invariance and the presence of critical
algebraic relaxation even down to $\tau=\infty$ one might expect
ageing to characterize the late time behavior of two-times functions
\cite{calabrese_ageing_2006}. Such behavior has been
predicted to appear after quenches to
critical points both in closed \cite{gambassi_2011,sciolla_2013,ciocchetta_2015} and open \cite{schmalian_2014}
quantum sytems.
In order to explore this possiblity we employ the fluctuation-dissipation ratio \cite{calabrese_ageing_2006}
$\chi_\mathcal{O}(t_1,t_2)= (G^R_\mathcal{O}(t_1,t_2) -G^A_\mathcal{O}(t_1,t_2))/\partial_{t_1}G^K_\mathcal{O}(t_1,t_2)$
with $t_1<t_2$,
which allows to address possible violations of detailed balance and
define effective temperatures for non-equilibrium systems, where the
fluctuation-dissipation theorem cannot be relied on.
In $\chi_\mathcal{O}(t_1,t_2)$ the index $\mathcal{O}$ means that the quotient is to be taken between expectation values corresponding to the most highly occupied eigenvector of some operator $\mathcal{O}$.
The limit 
$
\lim_{t_1\to\infty}\lim_{t_2\to\infty}
\chi_\mathcal{O}(t_1,t_2)\equiv 1/T_{\rm eff}
$
defines an effective temperature.
In systems exhibiting ageing after a quench to the critical point the
equilibration time diverges. As a consequence, the effective
temperature defined through the above limit will not be equal to the
value of the effective temperature obtained directly from the steady state,
even if the system is in contact with a thermal reservoir. 
Using our late-time GFs \eqref{eq:GF_longtime} it is easy to see that the fluctuation-dissipation ratio is independent of the relative time:
\begin{align}
\label{eq_chi_SCHF}
\chi(t_\text{rel},\tau)=\frac{1}{T_\text{eff}}\frac{1}{1+\frac{1}{\lambda_\text{kin}\kappa_\text{qp}\tau^2}}.
\end{align}
Therefore, for absolute times larger than the equilibration scale 
\begin{align}
\label{eq:equil_time}
\tau_\text{eq}=1/\sqrt{\lambda_\text{kin}\kappa_\text{qp}}
\end{align}
it relaxes to the inverse effective temperature $T_\text{eff}$.
Since at $g_c(N)$ the quasiparticle lifetime remains finite
$1/\kappa_{\rm qp}<\infty$, the equilibration scale $\tau_{\rm eq}$ is
also finite and thus no ageing takes place. $1/\kappa_{\rm qp}<\infty$ also implies that
the initial-slip exponent $\theta$ describing the $(t_2/t_1)^\theta$
scaling of two-times functions \cite{calabrese_ageing_2006} is
irrelevant, since the dynamics is exponential in the relative-time
direction (see Eq.~\eqref{eq:GF_longtime}).
However, due to the driven-dissipative nature of our system, the
steady state is not in global equilibrium, implying that the
effective temperature obtained from \eqref{eq_chi_SCHF} depends in general 
on the particular degree of freedom considered, consistent with what was
found in \cite{dallatorre_2013} by extracting $T_\text{eff}$ directly
from the steady state (see also Appendix \ref{app:F}).

\section{Predicitons for the experiment} 
The dynamical phase transition of the open Dicke
model has been investigated in recent quench experiments performed with a Bose-Einstein condensate (BEC) in an optical cavity
\cite{hemmerich_2015}. 
We expect our predictions to be observable in response and
correlation functions of the cavity output, once the wait-time
$\tau_{\rm w}$ after the quench satisfies $\tau_{\rm w}\gtrsim
\tau_{\rm alg}\sim 1/\kappa_{\rm qp}\sim N^{1/2}$ (see Section \ref{sec:results}).
Generically, the smallest value of $\tau_{\rm alg}$ is reached when
$\omega_z$ (corresponding to the recoil frequency $\omega_{\rm
  rec}\sim\text{KHz}$ in the BEC experiments) is of the same order of
$\kappa$. This can be seen by comparing the value of $\kappa_{\rm qp}$
for different values of $\kappa=2,1,0.2$ shown in Fig.~\ref{fig:kappacrossing}
and Fig.~\ref{fig:kappa02}, at a given $\omega_z=2.1$. The largest
$\kappa_{\rm qp}$ is reached indeed for $\kappa\simeq\omega_z$, while
for even larger $\kappa$ (not shown) the quasiparticle damping
decreases.
For instance, in the experimental setup of \cite{hemmerich_2015} the
cavity is very good: $\kappa\simeq\omega_{\rm rec}$, so that $\kappa_{\rm
  qp}\sim \kappa N^{-1/2}$
that is $\tau_{\rm w}\gtrsim \text{ms}\times (10^5)^{1/2}\sim 300 {\rm
  ms}$. While this is below typical BEC-lifetimes,
it is currently not achieved in the experiments \cite{hemmerich_2015}, but in
principle possible in the new-generation setups. 

While the measurement of response functions require
cavity probe-transmission experiments \cite{carmichael_2007,dallatorre_2013}, the
behavior of the correlation function in Fig.~\ref{fig:late-time_dyn}
will be directly observable from the cavity output intensity.

\section{Conclusions} 
We have shown that the dynamics following a
quench close to a critical point in an open quantum many-body system
can depend crucially on the competition between external and intrinsic
relaxation processes, the former due to drive and dissipation, the
latter due to integrability breaking through quasiparticle interactions.
In particular, we demonstrated a novel scenario involving a dynamical
phase transition where critical
algebraic relaxation is not accompanied by ageing, due to the finite
lifetime of quasiparticles.
The simplicity and paradigmatic character of the model considered
allowed for a detailed understanding of the phenomena and should imply
a broader relevance of our results.

\acknowledgements
We thank Philipp Strack for stimulating
discussions and feedback, especially in the intial stage of this work.
We also thank Alessio Chiocchetta for fruitful discussions. We are very grateful to
Wilhelm Zwerger for careful reading of the manuscript. 
F.P. is supported by the APART program of the Austrian Academy of Science.

 \appendix

\section{Keldysh formulation of the self-consistent Hartree-Fock theory}
\label{app:keldysh}

The open Dicke model introduced in the main text has already been
formulated within the Keldysh functional-integral framework \cite{dallatorre_2013}, thus we
only briefly discuss the main features here. We then describe the
self-consistent Hartree-Fock (SCHF) theory we employ.

Starting from the master equation, the functional-integral formulation
of the action is achieved by replacing the operators acting left(right)
of the density matrix with complex fields with a subscript
\textquotedblleft$+$\textquotedblright(\textquotedblleft$-$\textquotedblright). Calculating expectation values by a time-evolution along the Keldysh contour, the \textquotedblleft$+$\textquotedblright operators act while the system evolves forward in time, while the \textquotedblleft$-$\textquotedblright operators act on the backward branch.
It is easier to obtain
physical insight by rotating to \textquotedblleft
classical\textquotedblright\; $a_{cl}= \frac{1}{\sqrt{2}}\left(a_+ +
  a_-\right)$ and \textquotedblleft quantum\textquotedblright\;fields
$a_{q} = \frac{1}{\sqrt{2}}\left(a_+ - a_-\right)$, that deserve their name because only \textquotedblleft
classical\textquotedblright\; fields can propagate on-shell or have a
finite expectation value \cite{kamenev_book}, whereas \textquotedblleft
quantum\textquotedblright\; fields encode the (potentially correlated) statistical noise in an equivalent Langevin formulation.
Due to the loss of particle number conservation it is convenient to symmetrize the action through the identification of terms between advanced and retarded contributions.
The symmetrized action of $\hat{H}_0$ in the absence of coherent fields then reads \cite{dallatorre_2013}
\begin{align}\label{action}
S_0=\int \frac{d\omega}{2 \pi} V^\dagger(\omega)
\begin{pmatrix}
0 && \left[G^A_0\right]^{-1}(\omega)\\
\left[G^R_0\right]^{-1}(\omega) && D^K(\omega)\\
\end{pmatrix}
V(\omega)
\end{align}
because retarded and advanced Green's functions interchange under
$\omega\rightarrow-\omega$. The bare inverse GFs
$\left[G^R_0\right]^{-1}(\omega)$ and $D_0^K$ are given by
\begin{align}\label{eq:invGr}
&\left[G^R_0\right]^{-1}(\omega) =\nonumber\\
&\begin{pmatrix}
\omega-\omega_0+i \kappa && 0 && -g&&-g\\
0&& -\omega-\omega_0 - i \kappa && -g && -g\\
-g&&-g&&\omega-\omega_z && 0\\
-g&&-g&&0&&-\omega-\omega_z\\
\end{pmatrix}\nonumber\\
&\text{\;and\;\;}\nonumber\\
&D_0^K=\nonumber\\
&\begin{pmatrix}
2 i \kappa &&0&&0&&0\\
0&&2 i \kappa &&0&&0\\
0&&0&&0&&0\\
0&&0&&0&&0\\
\end{pmatrix}.
\end{align}
The verbose notation with the eight-component field
\begin{align}
V(\omega)=\left(\begin{array}{c}
a_{cl}(\omega)\\
a_{cl}^*(-\omega)\\
b_{cl}(\omega)\\
b_{cl}^*(-\omega)\\
a_{q}(\omega)\\
a_{q}^*(-\omega)\\
b_{q}(\omega)\\
b_{q}^*(-\omega)
\end{array}\right).
\end{align}
is necessary, since each -- Keldysh ($cl,q$) and Nambu ($\omega,-\omega$) structure -- double the number of fields compared to the quantum mechanical representation.

For $N<\infty$, the terms of the quartic interaction Hamiltonian
$\hat{H}'$ in Eq.~(\ref{eq:dm_hp}) have to be added to the action in \eqref{action}. Considering the possibility of interactions on the forward and the backward branch of the Keldysh contour, the corresponding part of the action reads
\begin{widetext}
\begin{align}
\label{eq:Sint}
\begin{split}
S_{\text{int}}&=\frac{g}{4 N}
\int\frac{d\omega}{2\pi}\bigg[\left[\left(a_{cl}+a_{cl}^*\right)\circ\left(b_{q}+b_{q}^*\right)+\left(a_{q}
+a_{q}^*\right) \circ\left(b_{cl}+b_{cl}^*\right)\right]\circ\left[b_{cl}^*b_{cl}+b_{q}^*b_q
\right]\\
&+\left[\left(a_{cl}+a_{cl}^*\right) \circ\left(b_{cl}+b_{cl}^*\right)+
\left(a_{q}+a_{q}^*\right) \circ\left(b_{q}+b_{q}^*\right)\right]\circ\left[b_{cl}^*b_q+
b_q^*b_{cl}\right]\bigg](\omega),
\end{split}
\end{align}
\end{widetext}
where \textquotedblleft $\circ$\textquotedblright\;denotes the convolution in $\omega$ (normalized by $1/(2 \pi)$).\\

\begin{figure}[htp]
\includegraphics[width=.95\columnwidth]{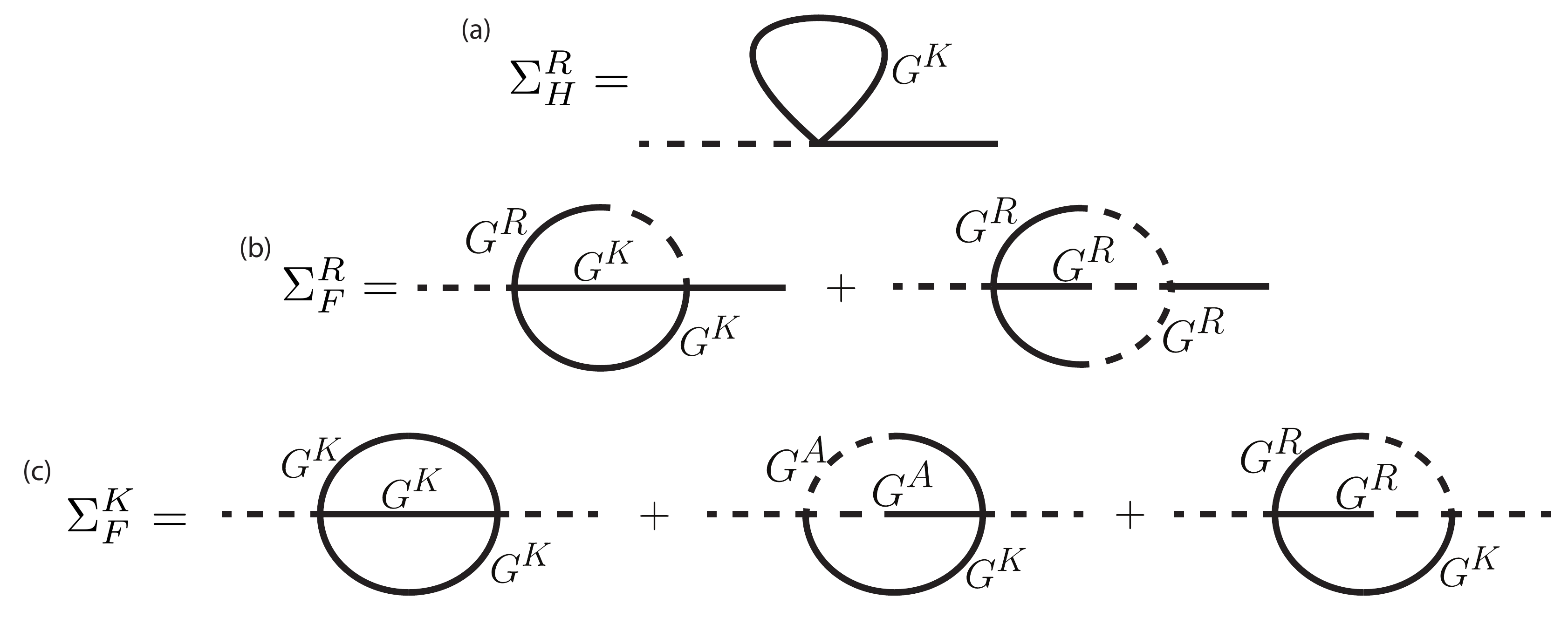}
\caption{Self-energy diagrams used in the SCHF calculations. The
  Hartree contribution (a) is only a real shift of the retarded
  component. The Fock contribution (b) and (c) is instead complex and
  frequency-dependent. 
  Here the solid(dashed) line correspond to a ``classical''(``quantum'') field attached to
  the vertex.}
\label{fig:diagrams}
\end{figure}
The results presented in the main text are obtained within a
self-consistent Hartree-Fock (SCHF) approximation, corresponding to
the selection of diagrams for the self-energies shown in
Fig.~\ref{fig:diagrams}. 
The self-consistent Hartree (SCH) approach has already been treated by Dalla Torre et al
in \cite{dallatorre_2013} for the steady state.
As we illustrate next, the inclusion of the Fock processes we perform
here is required to describe the effect of non-number-conserving quasiparticle collisions
breaking the integrability. These collisions are essential ingredients
in the steady state and late-time relaxation dynamics of the system
close to the superradiant transition.
Self-consistency is achieved by calculating the self-energies
$\Sigma^{(K,R)}$ as functionals of the dressed, rather than the bare Green's
functions: $\Sigma^{(K,R)}=\Sigma^{(K,R)}\!\!\left[G^R,G^K\right]$.
In order to highlight the novelties introduced by our SCHF approach,
we now briefly discuss the main features of the SCH theory.

Within the Hartree approximation only one skeleton diagram contributes to the self-energies. Furthermore, because $G^R(0)+G^A(0)=0$ only the retarded/advanced self-energy, given by the first diagram in Fig.~\ref{fig:diagrams}, is non-zero. The resulting frequency-independent self-consistence condition can be solved (mostly) analytically and predicts that both $\kappa_\text{qp}$ and the number of excitations in the steady state remain finite for all values of the coupling constant. Furthermore, it can be shown \footnote{J. Lang and
  F. Piazza, in preparation (2016)} that the steady state is
attractive under time-evolution for any coupling strength, implying
that no dynamical phase transition occurs on the self-consistent
one-loop level.

The theory becomes much more involved within the SCHF
approach we employ here. First of all, the Fock self-energy of
Fig.~\ref{fig:diagrams} is frequency-dependent as opposed to its
Hartree counterpart.
This enriches the problem by allowing for the inclusion of memory effects, which however play no significant role near the superradiant transition.
Additionally, the Keldysh component of the Fock self-energy
is nonzero and the retarded component has an imaginary part.
This implies that the inclusion of the Fock processes in our theory
allows us to describe relaxation through redistribution of energy via
collisions between quasiparticles.\\
Within this approximation there are two different subclasses of
diagrams: those involving only one "quantum" field and those with three
"quantum" fields. A bare scaling analysis of the model
(\ref{action}),(\ref{eq:Sint}) indicates that the latter are of higher order in $1/N$ compared to the more classical first subset of diagrams
\cite{dallatorre_2013}. Yet, in order to improve our quantitative results for
intermediate values of $N$ as well as for the phase transition, we keep
those diagrams. Independent of this, the self-consistent resummation
of two-loop diagrams cannot be performed analytically forcing us to
heavily rely on numerical methods for the calculation of quantitative
results. However, all the analytical expressions in the main text are
completely independent of the numerics, that can therefore -- as done
in figure 2 of the main text-- be used for independent confirmation.

\section{Time-integration of the coupled Dyson equations}
\label{app:numerics}

Within the SCHF approximation, we perform a time integration of the
coupled, nonlinear Dyson equations (5) and (6) for the Keldysh and retarded
GFs. We adopt an iteration procedure valid in the vicinity of the
steady state:
\begin{align}
\label{eq:iteration}
G^R(\tau+\delta \tau)=(1-c)G^R(\tau)+c \left(\left[G^R_0\right]^{-1}-\Sigma^R\left(G^R(\tau)\right)\right)^{-1}
\end{align}
where $c$ is the numerical update in our iteration, $\tau$ is the absolute time and we suppressed the dependence of the
GF on the relative time $t_{\rm rel}$. Here the subscript ss stands
for steady state. Since including the dynamics for the Keldysh component contributes
only further additive terms with the same global prefactors, we
simplify the expressions here to depend solely on the retarded Green's function.
We see how the approximate time-iteration (\ref{eq:iteration})
neglects memory effects involving time-integrals over the past, 
so that solutions of the form Eq. (7) of the main text
can be found, where the $t_{\rm rel}$-functional form depends only
parametrically on $\tau$ through $\kappa_{\rm qp}(\tau)$. The
$\tau$-dependence of the latter is of order $1/N$ and thus negligible.
The approximation involved in (\ref{eq:iteration}) relies on a
separation of timescales between the relative and absolute
time-evolution and is equivalent to taking the leading order in the Wigner
expansion of the convolutions between two-times functions \cite{Note1}:
\begin{align}
&A\circ B = \int dt_3 A(t_1,t_3) B(t_3,t_2)\nonumber\\ 
&\overset{WT}{\Leftrightarrow}  A(\tau,\omega)
e^{\frac{i}{2}\left(\overset{\leftarrow}{\partial_{\omega}}\overset{\rightarrow}{\partial_{\tau}}-\overset{\leftarrow}{\partial_{\tau}}\overset{\rightarrow}{\partial_{\omega}}\right)}B(\tau,\omega)\simeq A(\tau,\omega) B(\tau,\omega),
\end{align}
where we defined $f(t_1,t_2) \overset{WT}{\Leftrightarrow}
f(\tau,\omega)=\int dt_{\rm rel} e^{-i \omega t_{\rm rel}}
f(\tau-t_{\rm rel}/2,\tau+t_{\rm rel}/2)
$.
The required separation of timescales is achieved in our system in the
vicinity of the steady state and for a quench of $g$ close enough to
$g_c(N)$. As Fig.1 shows, in this regime
$\kappa_{\rm kin}$, setting the absolute timescale (see
Eq.(7)), is much smaller than $\kappa_{\rm
  qp}$. The latter, setting the relative timescale, remains indeed
finite at our dynamical phase transition. For the same reasons, our
numerical time-evolution is not applicable well inside the
weak-coupling regime (region I of Fig.1 in the main text), since there
$\kappa_{\rm kin}\simeq\kappa_{\rm qp}$.

\section{Role of the photon loss rate $\kappa$}
\label{app:kappa}

\begin{figure}[htp]
\includegraphics[width=\columnwidth]{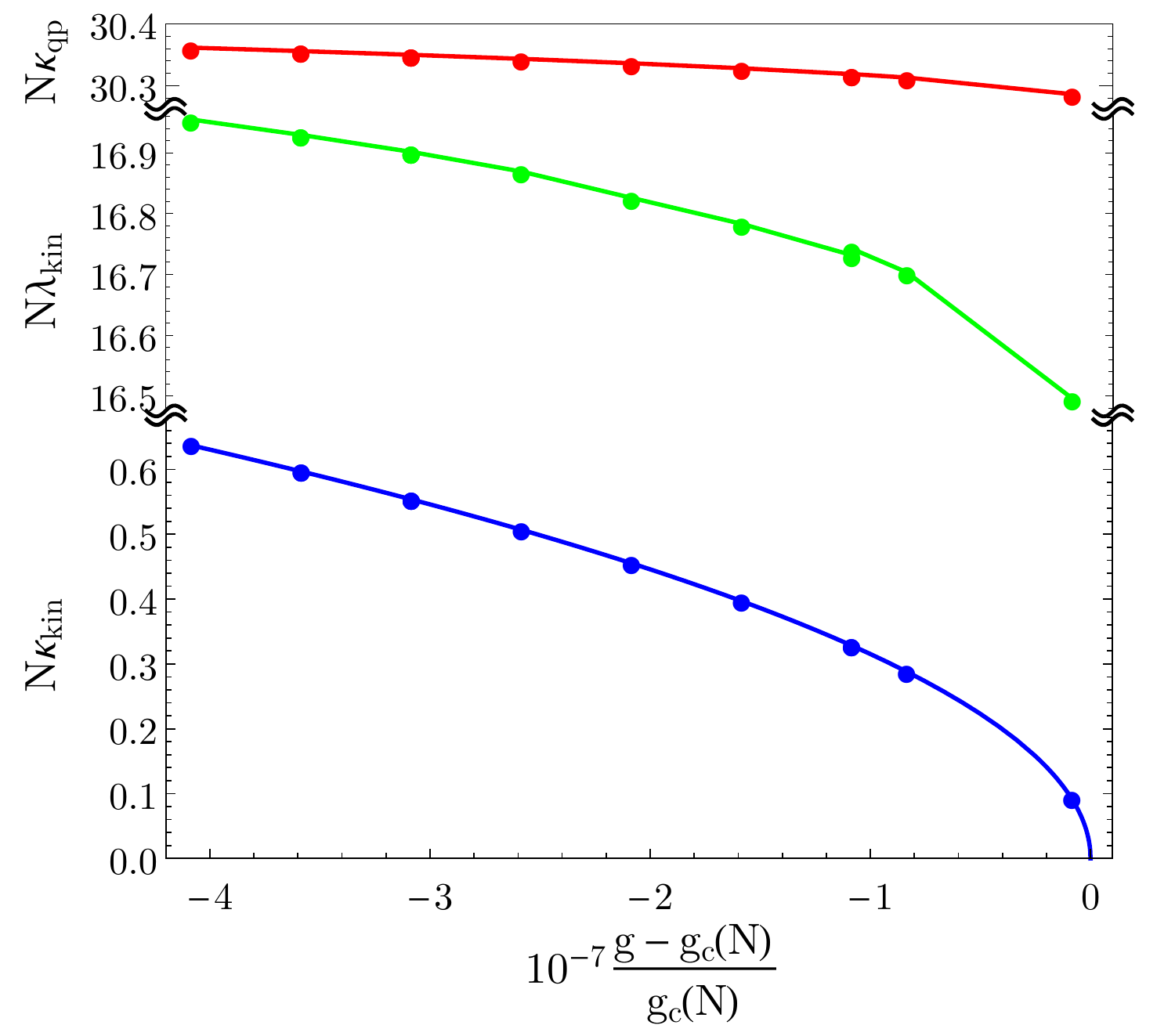}\\
\includegraphics[width=\columnwidth]{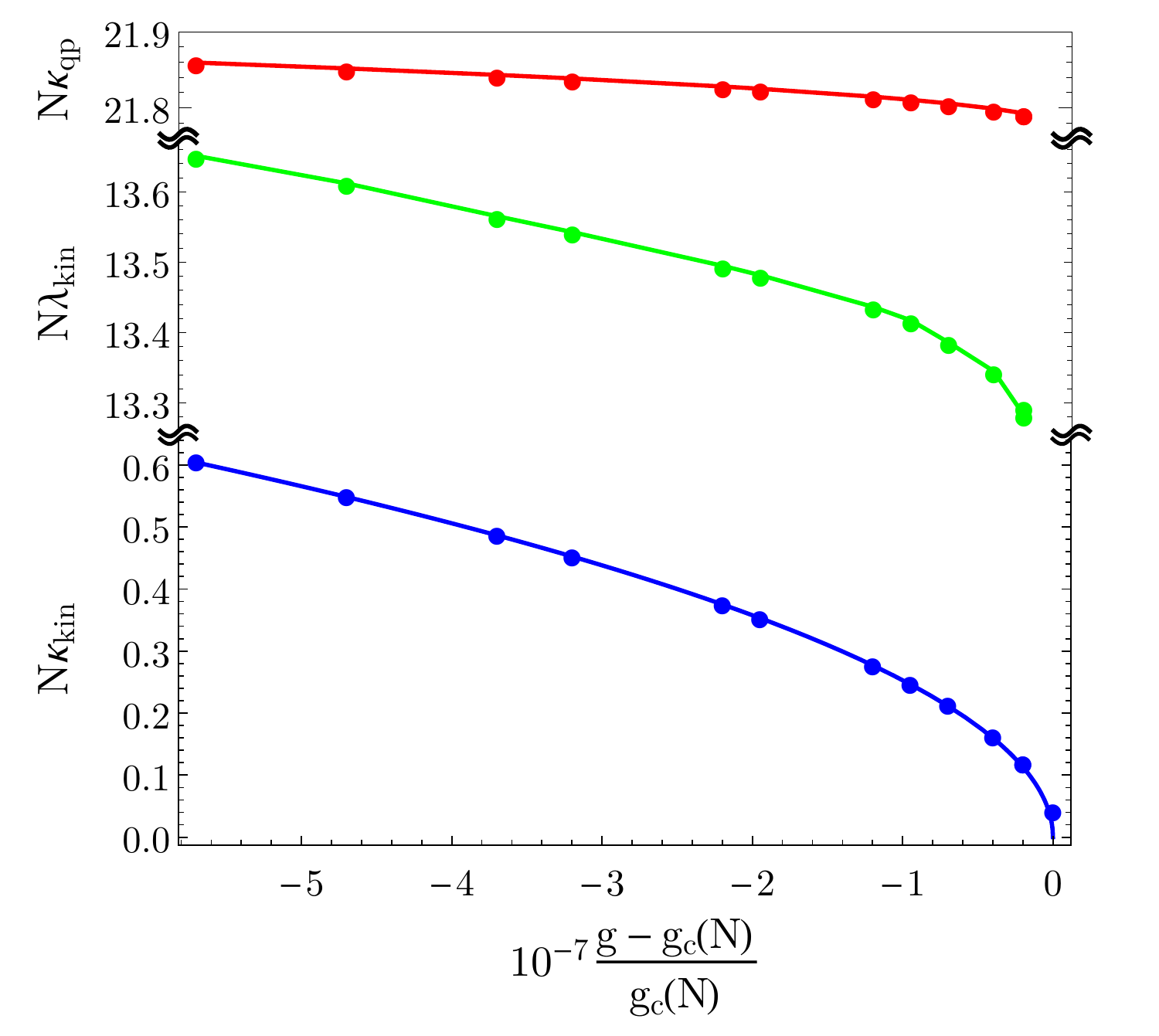}
\caption{Numerical values of the kinetic parameters
  $\kappa_\text{kin}$, $\lambda_\text{kin}$ as well as the inverse
  quasiparticle lifetime $\kappa_\text{qp}$. The parameters used are
  the same as in the main text, apart from $\kappa=1$ (upper panel)
  and $\kappa=0.2$ (lower panel), which result in $g_c\approx1.1420$ and $g_c\approx 1.0298$ respectively.}
\label{fig:kappa02}
\end{figure}

In this section we complement the results presented in the main text
by computing the dynamical parameters $\kappa_{\rm qp}$,
$\kappa_{\rm kin}$, $\lambda_{\rm kin}$ for smaller values of the
photon loss rate  $\kappa$. The goal is to illustrate the qualitative
behavior of the system in the isolated limit $\kappa\to 0$.
In Fig. \ref{fig:kappa02} we show the results for $\kappa=1$ and $\kappa=0.2$, to be compared
with Fig. 2 of the main text, computed for $\kappa=2$. Two main
observations emerge: i) since all the dynamical parameters ($\kappa_{\rm qp}$,
$\kappa_{\rm kin}$ , $\lambda_{\rm kin}$) decrease for
decreasing $\kappa$, the global timescale becomes slower; ii) since $\kappa_{\rm qp}$
and $\kappa_{\rm kin}$ become closer to one another, it becomes more
difficult (i.e. one has to tune the system even closer to $g_c(N)$) to
reach the dynamical critical regime where $\kappa_{\rm
  in}\ll\kappa_{\rm qp}$. Ultimately, in the $\kappa=0$ limit, we
expect $\kappa_{\rm kin}$ to become coupled to
$\kappa_{\rm qp}$, in the sense that the former cannot be made
arbitrarily small
compared to the latter, at any given $N$. The numerical computation leading to a
set of results as the one in Fig. \ref{fig:kappa02} is very demanding and becomes
more and more so a $\kappa\to 0$, since the global timescale becomes
slower and $\kappa_{\rm kin}\to\kappa_{\rm qp}$ (see previous section). Our approach is not applicable in
the case $\kappa=0$.

\section{The steady-state distribution function}
\label{app:F}

In this section we consider the steady-state of the coupled-Dyson
equations (5) and (6) of the main text. We will show how the
integrability-breaking through quasiparticle interactions leads to
equilibration. This intrinsic equilibration adds to the one induced by the coupling to the
external reservoir.

To this purpose, we consider the steady-state distribution function
$F(\omega)$ defined through
\begin{align}
\label{eq:dist_fun_def}
G^K(\omega) = G^R(\omega)\cdot F(\omega)-F(\omega)\cdot G^A(\omega).
\end{align}
The function $F(\omega)$ determines the link between response
and correlation functions and is therefore deeply connected with the
fluctuation-dissipation relations in the steady state \cite{kamenev_book}.
$F(\omega)$ describes the boundary conditions emergent in the steady
state for each degree of freedom of our system.
For instance, for a single bosonic degree of freedom in thermal
equilibrium with a reservoir at temperature $T$, the distribution function
$F(\omega)$ is simply $\coth(\omega/2T)$ \cite{kamenev_book}, while for a Markov reservoir
corresponding to the Lindblad operator (2) of the main text we have
$F(\omega)=1$, corresponding to a pure state \cite{dallatorre_2013}.
One can thus expect $F(\omega)$ to be sensitive to the different drive and
relaxation mechanisms, both external
and intrinsic to the system. 
\begin{figure}[htp]
\includegraphics[width=\columnwidth]{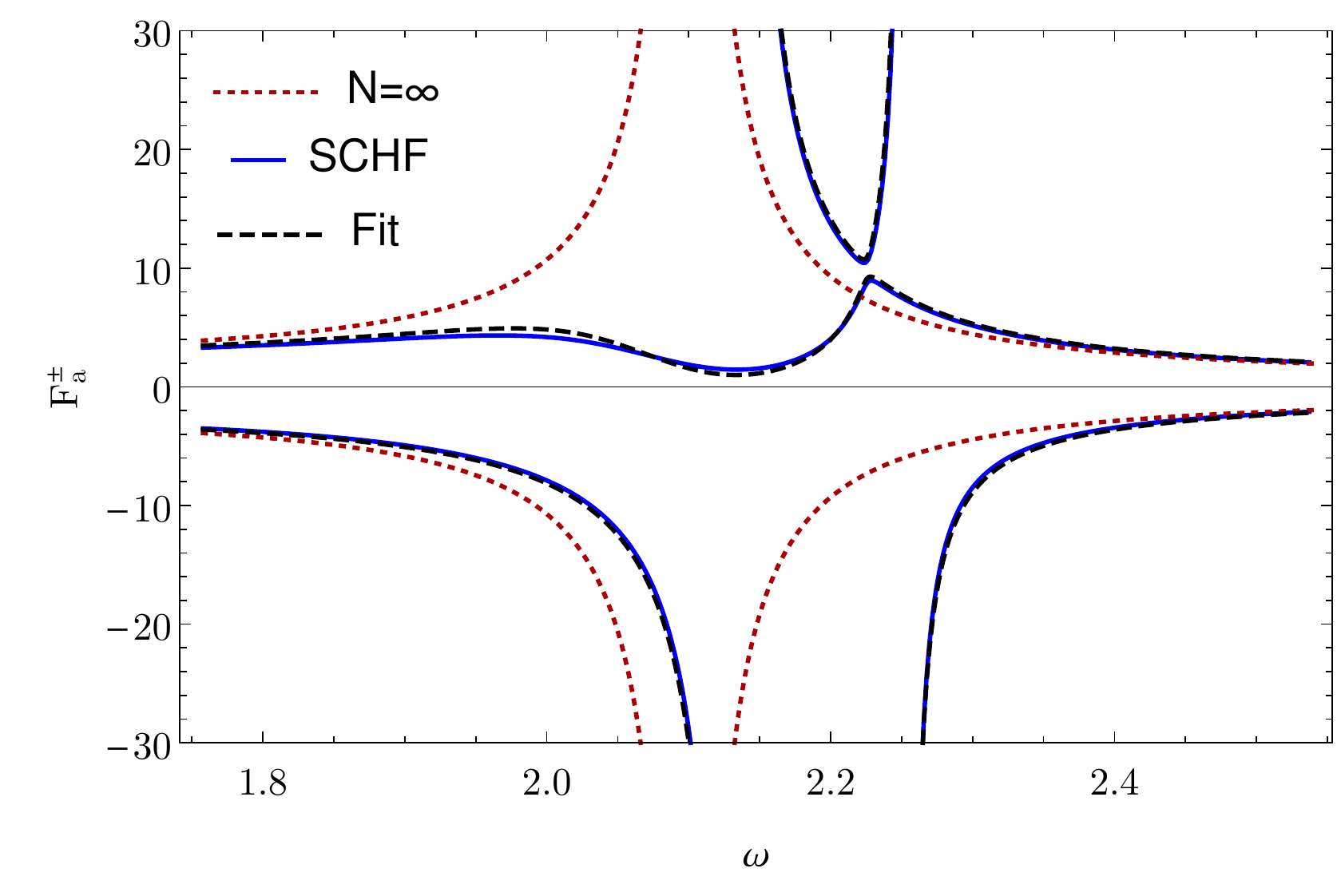}
\caption{Behavior of the two eigenvalues of the photonic distribution
  function. The red-dotted line corresponds to the integrable theory:
  $N=\infty$. The blue solid line is the result obtained from full
  SCHF theory, where qualitatively new features appear. These can be
  well described by an effective linearised theory Eq.~\eqref{eq:lin_theory} (black-dashed line)
  that shifts the atomic resonance in the complex plane and therefore
  contains only a single (complex) fit parameter $\kappa_b$. The parameters used are the same as in the main text, resulting in $\kappa_b\approx0.0335+0.0009i$.
}
\label{fig:fcomp}
\end{figure}

In order to analyze the distribution function for the photonic and
atomic degrees of freedom separately, we have first projected the full 4 by 4
Green's functions onto the respective 2 by 2 sectors. Within each
sector we then solved \eqref{eq:dist_fun_def} for $F_a(\omega)$ and $F_b (\omega)$,
respectively, where the subscript $a$ refers to the photonic and $b$
to the atomic sector \cite{Note1}. In Fig.~\ref{fig:fcomp}, we plot the eigenvalues $F^\pm_a(\omega)$ 
of $F_a(\omega)$, which due to the hermitian structure of $F(\omega)$ are purely real.

In our open Dicke model and in the integrable limit $N=\infty$, external driving is effectively present due to the
bilinear coupling between photonic and atomic degrees of freedom
which does not conserve the excitation number, while relaxation is
induced externally by photon losses. As discussed in
\cite{dallatorre_2013}, the corresponding distribution function for the photonic degree of freedom shows
singularities at zero frequency and at the bare atomic
resonance frequencies $\pm\omega_z$. These singularities appear on top of
the frequency-independent Markov background and result from the
effective drive via the atoms. These singularities are of thermal nature, behaving
like $T_{\rm eff}/\omega$, with the effective temperature emerging due
to the combination of the Markov reservoir and the driving. This temperature is
different for the photonic and atomic degrees of freedom, indicating
the violation of detailed balance arising from the fact that the whole
system is driven but dissipates only through the photons (see also
section \ref{sec:ageing}).

Within our SCHF approach, we are able to include the
equilibration mechanism intrinsic to the system, which is governed by the
integrability-breaking terms. In particular, as already discussed, the Fock processes allow to include the intrinsic
equilibration induced by quasiparticle collisions.
In the strong coupling regime: $(g_c(N)-g)/g_c(N)\lesssim N^{-1/2}$, this introduces large qualitative and quantitative changes in
$F(\omega)$, as illustrated in Fig.~\ref{fig:fcomp} for the photonic
degree of freedom. Here we compare the prediction of the integrable theory:
$N=\infty$ with our SCHF results. Apart from a shift of the
singularities from their bare value $\omega_z$, the important
qualitative change introduced by collisions is the splitting of these singularities
via an avoided crossing. This splitting of the singularities
at the (shifted) atomic resonances can be reproduced by adding to the integrable
theory a second Markov reservoir, this time for the atomic degree of
freedom. This corresponds to the steady-state of the following master equation
\begin{align}
\label{eq:lin_theory}
\partial_t \rho &= -i\left[\hat{H}_0,\rho\right]\nonumber\\&+\kappa \left(2 \hat{a}\rho \hat{a}^\dagger - \left\{\hat{a}^\dagger \hat{a}, \rho \right\}\right) +\kappa_b \left(2 \hat{b}\rho \hat{b}^\dagger - \left\{\hat{b}^\dagger \hat{b}, \rho \right\}\right).
\end{align}
where $\hat{H}_0$ indicates the integrable Hamiltonian of
Eq.(3).
By choosing the effective atomic dissipation $\kappa_b$ appropriately
(including the shift of the resonance frequency), we can simulate the
extent to which the quasiparticle collisions result in enhanced decay of atomic
excitations into multiple photons. This demonstrates how the
integrability-breaking leads to enhanced equilibration by creating
effectively a further bath for the the system.

While $F(\omega)$ contains a lot of information encoded in its
functional form, its measurement requires knowledge of both the spectral response
and the correlation functions. The former gives direct access to the retarded (and by complex conjugation the
advanced) Green's function while the latter directly corresponds to the Keldysh Green's function. 
Eq.~\eqref{eq:dist_fun_def} would then allow to compute the distribution function.

\bibliography{dicke_sib_biblio}

\begin{thebibliography}{86}%
\makeatletter
\providecommand \@ifxundefined [1]{%
 \@ifx{#1\undefined}
}%
\providecommand \@ifnum [1]{%
 \ifnum #1\expandafter \@firstoftwo
 \else \expandafter \@secondoftwo
 \fi
}%
\providecommand \@ifx [1]{%
 \ifx #1\expandafter \@firstoftwo
 \else \expandafter \@secondoftwo
 \fi
}%
\providecommand \natexlab [1]{#1}%
\providecommand \enquote  [1]{``#1''}%
\providecommand \bibnamefont  [1]{#1}%
\providecommand \bibfnamefont [1]{#1}%
\providecommand \citenamefont [1]{#1}%
\providecommand \href@noop [0]{\@secondoftwo}%
\providecommand \href [0]{\begingroup \@sanitize@url \@href}%
\providecommand \@href[1]{\@@startlink{#1}\@@href}%
\providecommand \@@href[1]{\endgroup#1\@@endlink}%
\providecommand \@sanitize@url [0]{\catcode `\\12\catcode `\$12\catcode
  `\&12\catcode `\#12\catcode `\^12\catcode `\_12\catcode `\%12\relax}%
\providecommand \@@startlink[1]{}%
\providecommand \@@endlink[0]{}%
\providecommand \url  [0]{\begingroup\@sanitize@url \@url }%
\providecommand \@url [1]{\endgroup\@href {#1}{\urlprefix }}%
\providecommand \urlprefix  [0]{URL }%
\providecommand \Eprint [0]{\href }%
\@ifxundefined \urlstyle {%
  \providecommand \doi  [0]{\begingroup \@sanitize@url \@doi}%
  \providecommand \@doi [1]{\endgroup \@@startlink {\doibase
  #1}doi:\discretionary {}{}{}#1\@@endlink }%
}{%
  \providecommand \doi  [0]{doi:\discretionary{}{}{}\begingroup
  \urlstyle{rm}\Url }%
}%
\providecommand \doibase [0]{http://dx.doi.org/}%
\providecommand \Doi [0]{\begingroup \@sanitize@url \@Doi }%
\providecommand \@Doi  [1]{\endgroup\@@startlink{\doibase#1}\@@Doi}%
\providecommand \@@Doi [1]{#1\@@endlink}%
\providecommand \selectlanguage [0]{\@gobble}%
\providecommand \bibinfo  [0]{\@secondoftwo}%
\providecommand \bibfield  [0]{\@secondoftwo}%
\providecommand \translation [1]{[#1]}%
\providecommand \BibitemOpen [0]{}%
\providecommand \bibitemStop [0]{}%
\providecommand \bibitemNoStop [0]{.\EOS\space}%
\providecommand \EOS [0]{\spacefactor3000\relax}%
\providecommand \BibitemShut  [1]{\csname bibitem#1\endcsname}%
\bibitem [{\citenamefont {Polkovnikov}\ \emph {et~al.}(2011)\citenamefont
  {Polkovnikov}, \citenamefont {Sengupta}, \citenamefont {Silva},\ and\
  \citenamefont {Vengalattore}}]{polkov_rev_2011}%
  \BibitemOpen
  \bibfield  {author} {\bibinfo {author} {\bibfnamefont {A.}~\bibnamefont
  {Polkovnikov}}, \bibinfo {author} {\bibfnamefont {K.}~\bibnamefont
  {Sengupta}}, \bibinfo {author} {\bibfnamefont {A.}~\bibnamefont {Silva}}, \
  and\ \bibinfo {author} {\bibfnamefont {M.}~\bibnamefont {Vengalattore}},\
  }\Doi {10.1103/RevModPhys.83.863} {\bibfield  {journal} {\bibinfo  {journal}
  {Rev. Mod. Phys.},\ }\textbf {\bibinfo {volume} {83}},\ \bibinfo {pages}
  {863} (\bibinfo {year} {2011})}\BibitemShut {NoStop}%
\bibitem [{\citenamefont {Eisert}\ \emph {et~al.}(2015)\citenamefont {Eisert},
  \citenamefont {Friesdorf},\ and\ \citenamefont
  {Gogolin}}]{gogolin_2015_nature}%
  \BibitemOpen
  \bibfield  {author} {\bibinfo {author} {\bibfnamefont {J.}~\bibnamefont
  {Eisert}}, \bibinfo {author} {\bibfnamefont {M.}~\bibnamefont {Friesdorf}}, \
  and\ \bibinfo {author} {\bibfnamefont {C.}~\bibnamefont {Gogolin}},\ }\href
  {http://dx.doi.org/10.1038/nphys3215} {\bibfield  {journal} {\bibinfo
  {journal} {Nat Phys},\ }\textbf {\bibinfo {volume} {11}},\ \bibinfo {pages}
  {124} (\bibinfo {year} {2015})}\BibitemShut {NoStop}%
\bibitem [{\citenamefont {Diehl}\ \emph {et~al.}(2008)\citenamefont {Diehl},
  \citenamefont {Micheli}, \citenamefont {Kantian}, \citenamefont {Kraus},
  \citenamefont {B{\"u}chler},\ and\ \citenamefont
  {Zoller}}]{diehl2008quantum}%
  \BibitemOpen
  \bibfield  {author} {\bibinfo {author} {\bibfnamefont {S.}~\bibnamefont
  {Diehl}}, \bibinfo {author} {\bibfnamefont {A.}~\bibnamefont {Micheli}},
  \bibinfo {author} {\bibfnamefont {A.}~\bibnamefont {Kantian}}, \bibinfo
  {author} {\bibfnamefont {B.}~\bibnamefont {Kraus}}, \bibinfo {author}
  {\bibfnamefont {H.}~\bibnamefont {B{\"u}chler}}, \ and\ \bibinfo {author}
  {\bibfnamefont {P.}~\bibnamefont {Zoller}},\ }\href@noop {} {\bibfield
  {journal} {\bibinfo  {journal} {Nature Physics},\ }\textbf {\bibinfo {volume}
  {4}},\ \bibinfo {pages} {878} (\bibinfo {year} {2008})}\BibitemShut {NoStop}%
\bibitem [{\citenamefont {Hohenberg}\ and\ \citenamefont
  {Halperin}(1977)}]{hohenberg1977theory}%
  \BibitemOpen
  \bibfield  {author} {\bibinfo {author} {\bibfnamefont {P.~C.}\ \bibnamefont
  {Hohenberg}}\ and\ \bibinfo {author} {\bibfnamefont {B.~I.}\ \bibnamefont
  {Halperin}},\ }\href@noop {} {\bibfield  {journal} {\bibinfo  {journal}
  {Reviews of Modern Physics},\ }\textbf {\bibinfo {volume} {49}},\ \bibinfo
  {pages} {435} (\bibinfo {year} {1977})}\BibitemShut {NoStop}%
\bibitem [{\citenamefont {Calabrese}\ and\ \citenamefont
  {Gambassi}(2005)}]{calabrese_ageing_2006}%
  \BibitemOpen
  \bibfield  {author} {\bibinfo {author} {\bibfnamefont {P.}~\bibnamefont
  {Calabrese}}\ and\ \bibinfo {author} {\bibfnamefont {A.}~\bibnamefont
  {Gambassi}},\ }\href {http://stacks.iop.org/0305-4470/38/i=18/a=R01}
  {\bibfield  {journal} {\bibinfo  {journal} {Journal of Physics A:
  Mathematical and General},\ }\textbf {\bibinfo {volume} {38}},\ \bibinfo
  {pages} {R133} (\bibinfo {year} {2005})}\BibitemShut {NoStop}%
\bibitem [{\citenamefont {Mitra}\ \emph {et~al.}(2006)\citenamefont {Mitra},
  \citenamefont {Takei}, \citenamefont {Kim},\ and\ \citenamefont
  {Millis}}]{millis_2006}%
  \BibitemOpen
  \bibfield  {author} {\bibinfo {author} {\bibfnamefont {A.}~\bibnamefont
  {Mitra}}, \bibinfo {author} {\bibfnamefont {S.}~\bibnamefont {Takei}},
  \bibinfo {author} {\bibfnamefont {Y.~B.}\ \bibnamefont {Kim}}, \ and\
  \bibinfo {author} {\bibfnamefont {A.~J.}\ \bibnamefont {Millis}},\ }\Doi
  {10.1103/PhysRevLett.97.236808} {\bibfield  {journal} {\bibinfo  {journal}
  {Phys. Rev. Lett.},\ }\textbf {\bibinfo {volume} {97}},\ \bibinfo {pages}
  {236808} (\bibinfo {year} {2006})}\BibitemShut {NoStop}%
\bibitem [{\citenamefont {Patan\`e}\ \emph {et~al.}(2008)\citenamefont
  {Patan\`e}, \citenamefont {Silva}, \citenamefont {Amico}, \citenamefont
  {Fazio},\ and\ \citenamefont {Santoro}}]{fazio_2008}%
  \BibitemOpen
  \bibfield  {author} {\bibinfo {author} {\bibfnamefont {D.}~\bibnamefont
  {Patan\`e}}, \bibinfo {author} {\bibfnamefont {A.}~\bibnamefont {Silva}},
  \bibinfo {author} {\bibfnamefont {L.}~\bibnamefont {Amico}}, \bibinfo
  {author} {\bibfnamefont {R.}~\bibnamefont {Fazio}}, \ and\ \bibinfo {author}
  {\bibfnamefont {G.~E.}\ \bibnamefont {Santoro}},\ }\Doi
  {10.1103/PhysRevLett.101.175701} {\bibfield  {journal} {\bibinfo  {journal}
  {Phys. Rev. Lett.},\ }\textbf {\bibinfo {volume} {101}},\ \bibinfo {pages}
  {175701} (\bibinfo {year} {2008})}\BibitemShut {NoStop}%
\bibitem [{\citenamefont {Diehl}\ \emph {et~al.}(2010)\citenamefont {Diehl},
  \citenamefont {Tomadin}, \citenamefont {Micheli}, \citenamefont {Fazio},\
  and\ \citenamefont {Zoller}}]{diehl_2010}%
  \BibitemOpen
  \bibfield  {author} {\bibinfo {author} {\bibfnamefont {S.}~\bibnamefont
  {Diehl}}, \bibinfo {author} {\bibfnamefont {A.}~\bibnamefont {Tomadin}},
  \bibinfo {author} {\bibfnamefont {A.}~\bibnamefont {Micheli}}, \bibinfo
  {author} {\bibfnamefont {R.}~\bibnamefont {Fazio}}, \ and\ \bibinfo {author}
  {\bibfnamefont {P.}~\bibnamefont {Zoller}},\ }\Doi
  {10.1103/PhysRevLett.105.015702} {\bibfield  {journal} {\bibinfo  {journal}
  {Phys. Rev. Lett.},\ }\textbf {\bibinfo {volume} {105}},\ \bibinfo {pages}
  {015702} (\bibinfo {year} {2010})}\BibitemShut {NoStop}%
\bibitem [{\citenamefont {Dalla~Torre}\ \emph {et~al.}(2010)\citenamefont
  {Dalla~Torre}, \citenamefont {Demler}, \citenamefont {Giamarchi},\ and\
  \citenamefont {Altman}}]{dallatorre_2010}%
  \BibitemOpen
  \bibfield  {author} {\bibinfo {author} {\bibfnamefont {E.~G.}\ \bibnamefont
  {Dalla~Torre}}, \bibinfo {author} {\bibfnamefont {E.}~\bibnamefont {Demler}},
  \bibinfo {author} {\bibfnamefont {T.}~\bibnamefont {Giamarchi}}, \ and\
  \bibinfo {author} {\bibfnamefont {E.}~\bibnamefont {Altman}},\ }\href
  {http://dx.doi.org/10.1038/nphys1754} {\bibfield  {journal} {\bibinfo
  {journal} {Nat Phys},\ }\textbf {\bibinfo {volume} {6}},\ \bibinfo {pages}
  {806} (\bibinfo {year} {2010})}\BibitemShut {NoStop}%
\bibitem [{\citenamefont {Klaers}\ \emph {et~al.}(2010)\citenamefont {Klaers},
  \citenamefont {Schmitt}, \citenamefont {Vewinger},\ and\ \citenamefont
  {Weitz}}]{weitz_2010}%
  \BibitemOpen
  \bibfield  {author} {\bibinfo {author} {\bibfnamefont {J.}~\bibnamefont
  {Klaers}}, \bibinfo {author} {\bibfnamefont {J.}~\bibnamefont {Schmitt}},
  \bibinfo {author} {\bibfnamefont {F.}~\bibnamefont {Vewinger}}, \ and\
  \bibinfo {author} {\bibfnamefont {M.}~\bibnamefont {Weitz}},\ }\href
  {http://dx.doi.org/10.1038/nature09567} {\bibfield  {journal} {\bibinfo
  {journal} {Nature},\ }\textbf {\bibinfo {volume} {468}},\ \bibinfo {pages}
  {545} (\bibinfo {year} {2010})}\BibitemShut {NoStop}%
\bibitem [{\citenamefont {Kessler}\ \emph {et~al.}(2012)\citenamefont
  {Kessler}, \citenamefont {Giedke}, \citenamefont {Imamoglu}, \citenamefont
  {Yelin}, \citenamefont {Lukin},\ and\ \citenamefont {Cirac}}]{kessler_2012}%
  \BibitemOpen
  \bibfield  {author} {\bibinfo {author} {\bibfnamefont {E.~M.}\ \bibnamefont
  {Kessler}}, \bibinfo {author} {\bibfnamefont {G.}~\bibnamefont {Giedke}},
  \bibinfo {author} {\bibfnamefont {A.}~\bibnamefont {Imamoglu}}, \bibinfo
  {author} {\bibfnamefont {S.~F.}\ \bibnamefont {Yelin}}, \bibinfo {author}
  {\bibfnamefont {M.~D.}\ \bibnamefont {Lukin}}, \ and\ \bibinfo {author}
  {\bibfnamefont {J.~I.}\ \bibnamefont {Cirac}},\ }\Doi
  {10.1103/PhysRevA.86.012116} {\bibfield  {journal} {\bibinfo  {journal}
  {Phys. Rev. A},\ }\textbf {\bibinfo {volume} {86}},\ \bibinfo {pages}
  {012116} (\bibinfo {year} {2012})}\BibitemShut {NoStop}%
\bibitem [{\citenamefont {Kirton}\ and\ \citenamefont
  {Keeling}(2013)}]{kirton2013nonequilibrium}%
  \BibitemOpen
  \bibfield  {author} {\bibinfo {author} {\bibfnamefont {P.}~\bibnamefont
  {Kirton}}\ and\ \bibinfo {author} {\bibfnamefont {J.}~\bibnamefont
  {Keeling}},\ }\href@noop {} {\bibfield  {journal} {\bibinfo  {journal}
  {Physical review letters},\ }\textbf {\bibinfo {volume} {111}},\ \bibinfo
  {pages} {100404} (\bibinfo {year} {2013})}\BibitemShut {NoStop}%
\bibitem [{\citenamefont {Sieberer}\ \emph {et~al.}(2013)\citenamefont
  {Sieberer}, \citenamefont {Huber}, \citenamefont {Altman},\ and\
  \citenamefont {Diehl}}]{sieberer_2013}%
  \BibitemOpen
  \bibfield  {author} {\bibinfo {author} {\bibfnamefont {L.~M.}\ \bibnamefont
  {Sieberer}}, \bibinfo {author} {\bibfnamefont {S.~D.}\ \bibnamefont {Huber}},
  \bibinfo {author} {\bibfnamefont {E.}~\bibnamefont {Altman}}, \ and\ \bibinfo
  {author} {\bibfnamefont {S.}~\bibnamefont {Diehl}},\ }\Doi
  {10.1103/PhysRevLett.110.195301} {\bibfield  {journal} {\bibinfo  {journal}
  {Phys. Rev. Lett.},\ }\textbf {\bibinfo {volume} {110}},\ \bibinfo {pages}
  {195301} (\bibinfo {year} {2013})}\BibitemShut {NoStop}%
\bibitem [{\citenamefont {T\"auber}\ and\ \citenamefont
  {Diehl}(2014)}]{tauber_2014}%
  \BibitemOpen
  \bibfield  {author} {\bibinfo {author} {\bibfnamefont {U.~C.}\ \bibnamefont
  {T\"auber}}\ and\ \bibinfo {author} {\bibfnamefont {S.}~\bibnamefont
  {Diehl}},\ }\Doi {10.1103/PhysRevX.4.021010} {\bibfield  {journal} {\bibinfo
  {journal} {Phys. Rev. X},\ }\textbf {\bibinfo {volume} {4}},\ \bibinfo
  {pages} {021010} (\bibinfo {year} {2014})}\BibitemShut {NoStop}%
\bibitem [{\citenamefont {Bonnes}\ \emph {et~al.}(2014)\citenamefont {Bonnes},
  \citenamefont {Charrier},\ and\ \citenamefont {L\"auchli}}]{bonnes_2014}%
  \BibitemOpen
  \bibfield  {author} {\bibinfo {author} {\bibfnamefont {L.}~\bibnamefont
  {Bonnes}}, \bibinfo {author} {\bibfnamefont {D.}~\bibnamefont {Charrier}}, \
  and\ \bibinfo {author} {\bibfnamefont {A.~M.}\ \bibnamefont {L\"auchli}},\
  }\Doi {10.1103/PhysRevA.90.033612} {\bibfield  {journal} {\bibinfo  {journal}
  {Phys. Rev. A},\ }\textbf {\bibinfo {volume} {90}},\ \bibinfo {pages}
  {033612} (\bibinfo {year} {2014})}\BibitemShut {NoStop}%
\bibitem [{\citenamefont {Marcuzzi}\ \emph {et~al.}(2014)\citenamefont
  {Marcuzzi}, \citenamefont {Levi}, \citenamefont {Diehl}, \citenamefont
  {Garrahan},\ and\ \citenamefont {Lesanovsky}}]{lesanovski_rydberg_2014}%
  \BibitemOpen
  \bibfield  {author} {\bibinfo {author} {\bibfnamefont {M.}~\bibnamefont
  {Marcuzzi}}, \bibinfo {author} {\bibfnamefont {E.}~\bibnamefont {Levi}},
  \bibinfo {author} {\bibfnamefont {S.}~\bibnamefont {Diehl}}, \bibinfo
  {author} {\bibfnamefont {J.~P.}\ \bibnamefont {Garrahan}}, \ and\ \bibinfo
  {author} {\bibfnamefont {I.}~\bibnamefont {Lesanovsky}},\ }\Doi
  {10.1103/PhysRevLett.113.210401} {\bibfield  {journal} {\bibinfo  {journal}
  {Phys. Rev. Lett.},\ }\textbf {\bibinfo {volume} {113}},\ \bibinfo {pages}
  {210401} (\bibinfo {year} {2014})}\BibitemShut {NoStop}%
\bibitem [{\citenamefont {Hoening}\ \emph {et~al.}(2014)\citenamefont
  {Hoening}, \citenamefont {Abdussalam}, \citenamefont {Fleischhauer},\ and\
  \citenamefont {Pohl}}]{pohl_ryd_2014}%
  \BibitemOpen
  \bibfield  {author} {\bibinfo {author} {\bibfnamefont {M.}~\bibnamefont
  {Hoening}}, \bibinfo {author} {\bibfnamefont {W.}~\bibnamefont {Abdussalam}},
  \bibinfo {author} {\bibfnamefont {M.}~\bibnamefont {Fleischhauer}}, \ and\
  \bibinfo {author} {\bibfnamefont {T.}~\bibnamefont {Pohl}},\ }\Doi
  {10.1103/PhysRevA.90.021603} {\bibfield  {journal} {\bibinfo  {journal}
  {Phys. Rev. A},\ }\textbf {\bibinfo {volume} {90}},\ \bibinfo {pages}
  {021603} (\bibinfo {year} {2014})}\BibitemShut {NoStop}%
\bibitem [{\citenamefont {Medvedyeva}\ \emph {et~al.}(2015)\citenamefont
  {Medvedyeva}, \citenamefont {\ifmmode \check{C}\else
  \v{C}\fi{}ubrovi\ifmmode~\acute{c}\else \'{c}\fi{}},\ and\ \citenamefont
  {Kehrein}}]{kehrein_2015}%
  \BibitemOpen
  \bibfield  {author} {\bibinfo {author} {\bibfnamefont {M.~V.}\ \bibnamefont
  {Medvedyeva}}, \bibinfo {author} {\bibfnamefont {M.~T.}\ \bibnamefont
  {\ifmmode \check{C}\else \v{C}\fi{}ubrovi\ifmmode~\acute{c}\else
  \'{c}\fi{}}}, \ and\ \bibinfo {author} {\bibfnamefont {S.}~\bibnamefont
  {Kehrein}},\ }\Doi {10.1103/PhysRevB.91.205416} {\bibfield  {journal}
  {\bibinfo  {journal} {Phys. Rev. B},\ }\textbf {\bibinfo {volume} {91}},\
  \bibinfo {pages} {205416} (\bibinfo {year} {2015})}\BibitemShut {NoStop}%
\bibitem [{\citenamefont {{Labouvie}}\ \emph {et~al.}(2015)\citenamefont
  {{Labouvie}}, \citenamefont {{Santra}}, \citenamefont {{Heun}},\ and\
  \citenamefont {{Ott}}}]{ott_2015}%
  \BibitemOpen
  \bibfield  {author} {\bibinfo {author} {\bibfnamefont {R.}~\bibnamefont
  {{Labouvie}}}, \bibinfo {author} {\bibfnamefont {B.}~\bibnamefont
  {{Santra}}}, \bibinfo {author} {\bibfnamefont {S.}~\bibnamefont {{Heun}}}, \
  and\ \bibinfo {author} {\bibfnamefont {H.}~\bibnamefont {{Ott}}},\
  }\href@noop {} {\bibfield  {journal} {\bibinfo  {journal} {ArXiv e-prints}}
  (\bibinfo {year} {2015})},\ \Eprint {http://arxiv.org/abs/1507.05007}
  {arXiv:1507.05007 [quant-ph]} \BibitemShut {NoStop}%
\bibitem [{\citenamefont {Maghrebi}\ and\ \citenamefont
  {Gorshkov}(2015)}]{maghrebi_2015}%
  \BibitemOpen
  \bibfield  {author} {\bibinfo {author} {\bibfnamefont {M.~F.}\ \bibnamefont
  {Maghrebi}}\ and\ \bibinfo {author} {\bibfnamefont {A.~V.}\ \bibnamefont
  {Gorshkov}},\ }\href@noop {} {\bibfield  {journal} {\bibinfo  {journal}
  {arXiv preprint arXiv:1507.01939}} (\bibinfo {year} {2015})}\BibitemShut
  {NoStop}%
\bibitem [{\citenamefont {Marino}\ and\ \citenamefont
  {Diehl}(2016)}]{marino_2016}%
  \BibitemOpen
  \bibfield  {author} {\bibinfo {author} {\bibfnamefont {J.}~\bibnamefont
  {Marino}}\ and\ \bibinfo {author} {\bibfnamefont {S.}~\bibnamefont {Diehl}},\
  }\Doi {10.1103/PhysRevLett.116.070407} {\bibfield  {journal} {\bibinfo
  {journal} {Phys. Rev. Lett.},\ }\textbf {\bibinfo {volume} {116}},\ \bibinfo
  {pages} {070407} (\bibinfo {year} {2016})}\BibitemShut {NoStop}%
\bibitem [{\citenamefont {Marino}\ and\ \citenamefont
  {Silva}(2012)}]{marino_noisy_2012}%
  \BibitemOpen
  \bibfield  {author} {\bibinfo {author} {\bibfnamefont {J.}~\bibnamefont
  {Marino}}\ and\ \bibinfo {author} {\bibfnamefont {A.}~\bibnamefont {Silva}},\
  }\Doi {10.1103/PhysRevB.86.060408} {\bibfield  {journal} {\bibinfo  {journal}
  {Phys. Rev. B},\ }\textbf {\bibinfo {volume} {86}},\ \bibinfo {pages}
  {060408} (\bibinfo {year} {2012})}\BibitemShut {NoStop}%
\bibitem [{\citenamefont {Sedlmayr}\ \emph {et~al.}(2013)\citenamefont
  {Sedlmayr}, \citenamefont {Ren}, \citenamefont {Gebhard},\ and\ \citenamefont
  {Sirker}}]{sirker_2013}%
  \BibitemOpen
  \bibfield  {author} {\bibinfo {author} {\bibfnamefont {N.}~\bibnamefont
  {Sedlmayr}}, \bibinfo {author} {\bibfnamefont {J.}~\bibnamefont {Ren}},
  \bibinfo {author} {\bibfnamefont {F.}~\bibnamefont {Gebhard}}, \ and\
  \bibinfo {author} {\bibfnamefont {J.}~\bibnamefont {Sirker}},\ }\Doi
  {10.1103/PhysRevLett.110.100406} {\bibfield  {journal} {\bibinfo  {journal}
  {Phys. Rev. Lett.},\ }\textbf {\bibinfo {volume} {110}},\ \bibinfo {pages}
  {100406} (\bibinfo {year} {2013})}\BibitemShut {NoStop}%
\bibitem [{\citenamefont {Cai}\ and\ \citenamefont
  {Barthel}(2013)}]{barthel_2013}%
  \BibitemOpen
  \bibfield  {author} {\bibinfo {author} {\bibfnamefont {Z.}~\bibnamefont
  {Cai}}\ and\ \bibinfo {author} {\bibfnamefont {T.}~\bibnamefont {Barthel}},\
  }\Doi {10.1103/PhysRevLett.111.150403} {\bibfield  {journal} {\bibinfo
  {journal} {Phys. Rev. Lett.},\ }\textbf {\bibinfo {volume} {111}},\ \bibinfo
  {pages} {150403} (\bibinfo {year} {2013})}\BibitemShut {NoStop}%
\bibitem [{\citenamefont {Horstmann}\ \emph {et~al.}(2013)\citenamefont
  {Horstmann}, \citenamefont {Cirac},\ and\ \citenamefont
  {Giedke}}]{giedke_2013}%
  \BibitemOpen
  \bibfield  {author} {\bibinfo {author} {\bibfnamefont {B.}~\bibnamefont
  {Horstmann}}, \bibinfo {author} {\bibfnamefont {J.~I.}\ \bibnamefont
  {Cirac}}, \ and\ \bibinfo {author} {\bibfnamefont {G.}~\bibnamefont
  {Giedke}},\ }\Doi {10.1103/PhysRevA.87.012108} {\bibfield  {journal}
  {\bibinfo  {journal} {Phys. Rev. A},\ }\textbf {\bibinfo {volume} {87}},\
  \bibinfo {pages} {012108} (\bibinfo {year} {2013})}\BibitemShut {NoStop}%
\bibitem [{\citenamefont {Foss-Feig}\ \emph {et~al.}(2013)\citenamefont
  {Foss-Feig}, \citenamefont {Hazzard}, \citenamefont {Bollinger},\ and\
  \citenamefont {Rey}}]{rey_2013}%
  \BibitemOpen
  \bibfield  {author} {\bibinfo {author} {\bibfnamefont {M.}~\bibnamefont
  {Foss-Feig}}, \bibinfo {author} {\bibfnamefont {K.~R.~A.}\ \bibnamefont
  {Hazzard}}, \bibinfo {author} {\bibfnamefont {J.~J.}\ \bibnamefont
  {Bollinger}}, \ and\ \bibinfo {author} {\bibfnamefont {A.~M.}\ \bibnamefont
  {Rey}},\ }\Doi {10.1103/PhysRevA.87.042101} {\bibfield  {journal} {\bibinfo
  {journal} {Phys. Rev. A},\ }\textbf {\bibinfo {volume} {87}},\ \bibinfo
  {pages} {042101} (\bibinfo {year} {2013})}\BibitemShut {NoStop}%
\bibitem [{\citenamefont {Gagel}\ \emph {et~al.}(2014)\citenamefont {Gagel},
  \citenamefont {Orth},\ and\ \citenamefont {Schmalian}}]{schmalian_2014}%
  \BibitemOpen
  \bibfield  {author} {\bibinfo {author} {\bibfnamefont {P.}~\bibnamefont
  {Gagel}}, \bibinfo {author} {\bibfnamefont {P.~P.}\ \bibnamefont {Orth}}, \
  and\ \bibinfo {author} {\bibfnamefont {J.}~\bibnamefont {Schmalian}},\ }\Doi
  {10.1103/PhysRevLett.113.220401} {\bibfield  {journal} {\bibinfo  {journal}
  {Phys. Rev. Lett.},\ }\textbf {\bibinfo {volume} {113}},\ \bibinfo {pages}
  {220401} (\bibinfo {year} {2014})}\BibitemShut {NoStop}%
\bibitem [{\citenamefont {Piazza}\ and\ \citenamefont
  {Strack}(2014)}]{piazza_QKE}%
  \BibitemOpen
  \bibfield  {author} {\bibinfo {author} {\bibfnamefont {F.}~\bibnamefont
  {Piazza}}\ and\ \bibinfo {author} {\bibfnamefont {P.}~\bibnamefont
  {Strack}},\ }\Doi {10.1103/PhysRevA.90.043823} {\bibfield  {journal}
  {\bibinfo  {journal} {Phys. Rev. A},\ }\textbf {\bibinfo {volume} {90}},\
  \bibinfo {pages} {043823} (\bibinfo {year} {2014})}\BibitemShut {NoStop}%
\bibitem [{\citenamefont {Sch\"utz}\ and\ \citenamefont
  {Morigi}(2014)}]{schuetz_2014}%
  \BibitemOpen
  \bibfield  {author} {\bibinfo {author} {\bibfnamefont {S.}~\bibnamefont
  {Sch\"utz}}\ and\ \bibinfo {author} {\bibfnamefont {G.}~\bibnamefont
  {Morigi}},\ }\Doi {10.1103/PhysRevLett.113.203002} {\bibfield  {journal}
  {\bibinfo  {journal} {Phys. Rev. Lett.},\ }\textbf {\bibinfo {volume}
  {113}},\ \bibinfo {pages} {203002} (\bibinfo {year} {2014})}\BibitemShut
  {NoStop}%
\bibitem [{\citenamefont {Sch\"utz}\ \emph {et~al.}(2015)\citenamefont
  {Sch\"utz}, \citenamefont {J\"ager},\ and\ \citenamefont
  {Morigi}}]{schuetz_2015}%
  \BibitemOpen
  \bibfield  {author} {\bibinfo {author} {\bibfnamefont {S.}~\bibnamefont
  {Sch\"utz}}, \bibinfo {author} {\bibfnamefont {S.~B.}\ \bibnamefont
  {J\"ager}}, \ and\ \bibinfo {author} {\bibfnamefont {G.}~\bibnamefont
  {Morigi}},\ }\Doi {10.1103/PhysRevA.92.063808} {\bibfield  {journal}
  {\bibinfo  {journal} {Phys. Rev. A},\ }\textbf {\bibinfo {volume} {92}},\
  \bibinfo {pages} {063808} (\bibinfo {year} {2015})}\BibitemShut {NoStop}%
\bibitem [{\citenamefont {Sciolla}\ \emph {et~al.}(2015)\citenamefont
  {Sciolla}, \citenamefont {Poletti},\ and\ \citenamefont
  {Kollath}}]{kollath_2015}%
  \BibitemOpen
  \bibfield  {author} {\bibinfo {author} {\bibfnamefont {B.}~\bibnamefont
  {Sciolla}}, \bibinfo {author} {\bibfnamefont {D.}~\bibnamefont {Poletti}}, \
  and\ \bibinfo {author} {\bibfnamefont {C.}~\bibnamefont {Kollath}},\ }\Doi
  {10.1103/PhysRevLett.114.170401} {\bibfield  {journal} {\bibinfo  {journal}
  {Phys. Rev. Lett.},\ }\textbf {\bibinfo {volume} {114}},\ \bibinfo {pages}
  {170401} (\bibinfo {year} {2015})}\BibitemShut {NoStop}%
\bibitem [{\citenamefont {Buchhold}\ and\ \citenamefont
  {Diehl}(2015)}]{buchhold_lutt_2015}%
  \BibitemOpen
  \bibfield  {author} {\bibinfo {author} {\bibfnamefont {M.}~\bibnamefont
  {Buchhold}}\ and\ \bibinfo {author} {\bibfnamefont {S.}~\bibnamefont
  {Diehl}},\ }\Doi {10.1103/PhysRevA.92.013603} {\bibfield  {journal} {\bibinfo
   {journal} {Phys. Rev. A},\ }\textbf {\bibinfo {volume} {92}},\ \bibinfo
  {pages} {013603} (\bibinfo {year} {2015})}\BibitemShut {NoStop}%
\bibitem [{\citenamefont {Blatt}\ and\ \citenamefont
  {Roos}(2012)}]{blatt2012quantum}%
  \BibitemOpen
  \bibfield  {author} {\bibinfo {author} {\bibfnamefont {R.}~\bibnamefont
  {Blatt}}\ and\ \bibinfo {author} {\bibfnamefont {C.}~\bibnamefont {Roos}},\
  }\href@noop {} {\bibfield  {journal} {\bibinfo  {journal} {Nature Physics},\
  }\textbf {\bibinfo {volume} {8}},\ \bibinfo {pages} {277} (\bibinfo {year}
  {2012})}\BibitemShut {NoStop}%
\bibitem [{\citenamefont {Britton}\ \emph {et~al.}(2012)\citenamefont
  {Britton}, \citenamefont {Sawyer}, \citenamefont {Keith}, \citenamefont
  {Wang}, \citenamefont {Freericks}, \citenamefont {Uys}, \citenamefont
  {Biercuk},\ and\ \citenamefont {Bollinger}}]{britton2012engineered}%
  \BibitemOpen
  \bibfield  {author} {\bibinfo {author} {\bibfnamefont {J.~W.}\ \bibnamefont
  {Britton}}, \bibinfo {author} {\bibfnamefont {B.~C.}\ \bibnamefont {Sawyer}},
  \bibinfo {author} {\bibfnamefont {A.~C.}\ \bibnamefont {Keith}}, \bibinfo
  {author} {\bibfnamefont {C.-C.~J.}\ \bibnamefont {Wang}}, \bibinfo {author}
  {\bibfnamefont {J.~K.}\ \bibnamefont {Freericks}}, \bibinfo {author}
  {\bibfnamefont {H.}~\bibnamefont {Uys}}, \bibinfo {author} {\bibfnamefont
  {M.~J.}\ \bibnamefont {Biercuk}}, \ and\ \bibinfo {author} {\bibfnamefont
  {J.~J.}\ \bibnamefont {Bollinger}},\ }\href@noop {} {\bibfield  {journal}
  {\bibinfo  {journal} {Nature},\ }\textbf {\bibinfo {volume} {484}},\ \bibinfo
  {pages} {489} (\bibinfo {year} {2012})}\BibitemShut {NoStop}%
\bibitem [{\citenamefont {Carusotto}\ and\ \citenamefont
  {Ciuti}(2013)}]{carusotto_rev_2013}%
  \BibitemOpen
  \bibfield  {author} {\bibinfo {author} {\bibfnamefont {I.}~\bibnamefont
  {Carusotto}}\ and\ \bibinfo {author} {\bibfnamefont {C.}~\bibnamefont
  {Ciuti}},\ }\Doi {10.1103/RevModPhys.85.299} {\bibfield  {journal} {\bibinfo
  {journal} {Rev. Mod. Phys.},\ }\textbf {\bibinfo {volume} {85}},\ \bibinfo
  {pages} {299} (\bibinfo {year} {2013})}\BibitemShut {NoStop}%
\bibitem [{\citenamefont {Hartmann}\ \emph {et~al.}(2008)\citenamefont
  {Hartmann}, \citenamefont {Brandao},\ and\ \citenamefont
  {Plenio}}]{hartmann2008laser}%
  \BibitemOpen
  \bibfield  {author} {\bibinfo {author} {\bibfnamefont {M.~J.}\ \bibnamefont
  {Hartmann}}, \bibinfo {author} {\bibfnamefont {F.}~\bibnamefont {Brandao}}, \
  and\ \bibinfo {author} {\bibfnamefont {M.~B.}\ \bibnamefont {Plenio}},\
  }\href@noop {} {\bibfield  {journal} {\bibinfo  {journal} {Rev},\ }\textbf
  {\bibinfo {volume} {2}},\ \bibinfo {pages} {527} (\bibinfo {year}
  {2008})}\BibitemShut {NoStop}%
\bibitem [{\citenamefont {Houck}\ \emph {et~al.}(2012)\citenamefont {Houck},
  \citenamefont {T{\"u}reci},\ and\ \citenamefont {Koch}}]{houck2012chip}%
  \BibitemOpen
  \bibfield  {author} {\bibinfo {author} {\bibfnamefont {A.~A.}\ \bibnamefont
  {Houck}}, \bibinfo {author} {\bibfnamefont {H.~E.}\ \bibnamefont
  {T{\"u}reci}}, \ and\ \bibinfo {author} {\bibfnamefont {J.}~\bibnamefont
  {Koch}},\ }\href@noop {} {\bibfield  {journal} {\bibinfo  {journal} {Nature
  Physics},\ }\textbf {\bibinfo {volume} {8}},\ \bibinfo {pages} {292}
  (\bibinfo {year} {2012})}\BibitemShut {NoStop}%
\bibitem [{\citenamefont {Schmidt}\ and\ \citenamefont
  {Koch}(2013)}]{schmidt2013circuit}%
  \BibitemOpen
  \bibfield  {author} {\bibinfo {author} {\bibfnamefont {S.}~\bibnamefont
  {Schmidt}}\ and\ \bibinfo {author} {\bibfnamefont {J.}~\bibnamefont {Koch}},\
  }\href@noop {} {\bibfield  {journal} {\bibinfo  {journal} {Annalen der
  Physik},\ }\textbf {\bibinfo {volume} {525}},\ \bibinfo {pages} {395}
  (\bibinfo {year} {2013})}\BibitemShut {NoStop}%
\bibitem [{\citenamefont {Ludwig}\ and\ \citenamefont
  {Marquardt}(2013)}]{ludwig2013quantum}%
  \BibitemOpen
  \bibfield  {author} {\bibinfo {author} {\bibfnamefont {M.}~\bibnamefont
  {Ludwig}}\ and\ \bibinfo {author} {\bibfnamefont {F.}~\bibnamefont
  {Marquardt}},\ }\href@noop {} {\bibfield  {journal} {\bibinfo  {journal}
  {Physical review letters},\ }\textbf {\bibinfo {volume} {111}},\ \bibinfo
  {pages} {073603} (\bibinfo {year} {2013})}\BibitemShut {NoStop}%
\bibitem [{\citenamefont {Vetsch}\ \emph {et~al.}(2010)\citenamefont {Vetsch},
  \citenamefont {Reitz}, \citenamefont {Sagu\'e}, \citenamefont {Schmidt},
  \citenamefont {Dawkins},\ and\ \citenamefont
  {Rauschenbeutel}}]{rauschenb_2010}%
  \BibitemOpen
  \bibfield  {author} {\bibinfo {author} {\bibfnamefont {E.}~\bibnamefont
  {Vetsch}}, \bibinfo {author} {\bibfnamefont {D.}~\bibnamefont {Reitz}},
  \bibinfo {author} {\bibfnamefont {G.}~\bibnamefont {Sagu\'e}}, \bibinfo
  {author} {\bibfnamefont {R.}~\bibnamefont {Schmidt}}, \bibinfo {author}
  {\bibfnamefont {S.~T.}\ \bibnamefont {Dawkins}}, \ and\ \bibinfo {author}
  {\bibfnamefont {A.}~\bibnamefont {Rauschenbeutel}},\ }\Doi
  {10.1103/PhysRevLett.104.203603} {\bibfield  {journal} {\bibinfo  {journal}
  {Phys. Rev. Lett.},\ }\textbf {\bibinfo {volume} {104}},\ \bibinfo {pages}
  {203603} (\bibinfo {year} {2010})}\BibitemShut {NoStop}%
\bibitem [{\citenamefont {Ritsch}\ \emph {et~al.}(2013)\citenamefont {Ritsch},
  \citenamefont {Domokos}, \citenamefont {Brennecke},\ and\ \citenamefont
  {Esslinger}}]{cavity_rmp}%
  \BibitemOpen
  \bibfield  {author} {\bibinfo {author} {\bibfnamefont {H.}~\bibnamefont
  {Ritsch}}, \bibinfo {author} {\bibfnamefont {P.}~\bibnamefont {Domokos}},
  \bibinfo {author} {\bibfnamefont {F.}~\bibnamefont {Brennecke}}, \ and\
  \bibinfo {author} {\bibfnamefont {T.}~\bibnamefont {Esslinger}},\ }\Doi
  {10.1103/RevModPhys.85.553} {\bibfield  {journal} {\bibinfo  {journal} {Rev.
  Mod. Phys.},\ }\textbf {\bibinfo {volume} {85}},\ \bibinfo {pages} {553}
  (\bibinfo {year} {2013})}\BibitemShut {NoStop}%
\bibitem [{\citenamefont {Goban}\ \emph {et~al.}(2014)\citenamefont {Goban},
  \citenamefont {Hung}, \citenamefont {Yu}, \citenamefont {Hood}, \citenamefont
  {Muniz}, \citenamefont {Lee}, \citenamefont {Martin}, \citenamefont
  {McClung}, \citenamefont {Choi}, \citenamefont {Chang}, \citenamefont
  {Painter},\ and\ \citenamefont {Kimble}}]{kimble_2014_crystal}%
  \BibitemOpen
  \bibfield  {author} {\bibinfo {author} {\bibfnamefont {A.}~\bibnamefont
  {Goban}}, \bibinfo {author} {\bibfnamefont {C.~L.}\ \bibnamefont {Hung}},
  \bibinfo {author} {\bibfnamefont {S.~P.}\ \bibnamefont {Yu}}, \bibinfo
  {author} {\bibfnamefont {J.~D.}\ \bibnamefont {Hood}}, \bibinfo {author}
  {\bibfnamefont {J.~A.}\ \bibnamefont {Muniz}}, \bibinfo {author}
  {\bibfnamefont {J.~H.}\ \bibnamefont {Lee}}, \bibinfo {author} {\bibfnamefont
  {M.~J.}\ \bibnamefont {Martin}}, \bibinfo {author} {\bibfnamefont {A.~C.}\
  \bibnamefont {McClung}}, \bibinfo {author} {\bibfnamefont {K.~S.}\
  \bibnamefont {Choi}}, \bibinfo {author} {\bibfnamefont {D.~E.}\ \bibnamefont
  {Chang}}, \bibinfo {author} {\bibfnamefont {O.}~\bibnamefont {Painter}}, \
  and\ \bibinfo {author} {\bibfnamefont {H.~J.}\ \bibnamefont {Kimble}},\
  }\href {http://dx.doi.org/10.1038/ncomms4808} {\bibfield  {journal} {\bibinfo
   {journal} {Nat Commun},\ }\textbf {\bibinfo {volume} {5}} (\bibinfo {year}
  {2014})}\BibitemShut {NoStop}%
\bibitem [{\citenamefont {Douglas}\ \emph {et~al.}(2015)\citenamefont
  {Douglas}, \citenamefont {Habibian}, \citenamefont {Hung}, \citenamefont
  {Gorshkov}, \citenamefont {Kimble},\ and\ \citenamefont
  {Chang}}]{chang_many_body_2015}%
  \BibitemOpen
  \bibfield  {author} {\bibinfo {author} {\bibfnamefont {J.~S.}\ \bibnamefont
  {Douglas}}, \bibinfo {author} {\bibfnamefont {H.}~\bibnamefont {Habibian}},
  \bibinfo {author} {\bibfnamefont {C.~L.}\ \bibnamefont {Hung}}, \bibinfo
  {author} {\bibfnamefont {A.~V.}\ \bibnamefont {Gorshkov}}, \bibinfo {author}
  {\bibfnamefont {H.~J.}\ \bibnamefont {Kimble}}, \ and\ \bibinfo {author}
  {\bibfnamefont {D.~E.}\ \bibnamefont {Chang}},\ }\href
  {http://dx.doi.org/10.1038/nphoton.2015.57} {\bibfield  {journal} {\bibinfo
  {journal} {Nat Photon},\ }\textbf {\bibinfo {volume} {9}},\ \bibinfo {pages}
  {326} (\bibinfo {year} {2015})}\BibitemShut {NoStop}%
\bibitem [{\citenamefont {Dimer}\ \emph {et~al.}(2007)\citenamefont {Dimer},
  \citenamefont {Estienne}, \citenamefont {Parkins},\ and\ \citenamefont
  {Carmichael}}]{carmichael_2007}%
  \BibitemOpen
  \bibfield  {author} {\bibinfo {author} {\bibfnamefont {F.}~\bibnamefont
  {Dimer}}, \bibinfo {author} {\bibfnamefont {B.}~\bibnamefont {Estienne}},
  \bibinfo {author} {\bibfnamefont {A.~S.}\ \bibnamefont {Parkins}}, \ and\
  \bibinfo {author} {\bibfnamefont {H.~J.}\ \bibnamefont {Carmichael}},\ }\Doi
  {10.1103/PhysRevA.75.013804} {\bibfield  {journal} {\bibinfo  {journal}
  {Phys. Rev. A},\ }\textbf {\bibinfo {volume} {75}},\ \bibinfo {pages}
  {013804} (\bibinfo {year} {2007})}\BibitemShut {NoStop}%
\bibitem [{\citenamefont {Keeling}\ \emph {et~al.}(2010)\citenamefont
  {Keeling}, \citenamefont {Bhaseen},\ and\ \citenamefont
  {Simons}}]{simons_2010}%
  \BibitemOpen
  \bibfield  {author} {\bibinfo {author} {\bibfnamefont {J.}~\bibnamefont
  {Keeling}}, \bibinfo {author} {\bibfnamefont {M.~J.}\ \bibnamefont
  {Bhaseen}}, \ and\ \bibinfo {author} {\bibfnamefont {B.~D.}\ \bibnamefont
  {Simons}},\ }\Doi {10.1103/PhysRevLett.105.043001} {\bibfield  {journal}
  {\bibinfo  {journal} {Phys. Rev. Lett.},\ }\textbf {\bibinfo {volume}
  {105}},\ \bibinfo {pages} {043001} (\bibinfo {year} {2010})}\BibitemShut
  {NoStop}%
\bibitem [{\citenamefont {Bhaseen}\ \emph {et~al.}(2012)\citenamefont
  {Bhaseen}, \citenamefont {Mayoh}, \citenamefont {Simons},\ and\ \citenamefont
  {Keeling}}]{bhaseen_2012}%
  \BibitemOpen
  \bibfield  {author} {\bibinfo {author} {\bibfnamefont {M.~J.}\ \bibnamefont
  {Bhaseen}}, \bibinfo {author} {\bibfnamefont {J.}~\bibnamefont {Mayoh}},
  \bibinfo {author} {\bibfnamefont {B.~D.}\ \bibnamefont {Simons}}, \ and\
  \bibinfo {author} {\bibfnamefont {J.}~\bibnamefont {Keeling}},\ }\Doi
  {10.1103/PhysRevA.85.013817} {\bibfield  {journal} {\bibinfo  {journal}
  {Phys. Rev. A},\ }\textbf {\bibinfo {volume} {85}},\ \bibinfo {pages}
  {013817} (\bibinfo {year} {2012})}\BibitemShut {NoStop}%
\bibitem [{\citenamefont {\"Oztop}\ \emph {et~al.}(2012)\citenamefont
  {\"Oztop}, \citenamefont {Bordyuh}, \citenamefont {M\"ustecaplioglu},\ and\
  \citenamefont {T\"ureci}}]{tureci_2012}%
  \BibitemOpen
  \bibfield  {author} {\bibinfo {author} {\bibfnamefont {B.}~\bibnamefont
  {\"Oztop}}, \bibinfo {author} {\bibfnamefont {M.}~\bibnamefont {Bordyuh}},
  \bibinfo {author} {\bibfnamefont {O.~E.}\ \bibnamefont {M\"ustecaplioglu}}, \
  and\ \bibinfo {author} {\bibfnamefont {H.~E.}\ \bibnamefont {T\"ureci}},\
  }\href {http://stacks.iop.org/1367-2630/14/i=8/a=085011} {\bibfield
  {journal} {\bibinfo  {journal} {New Journal of Physics},\ }\textbf {\bibinfo
  {volume} {14}},\ \bibinfo {pages} {085011} (\bibinfo {year}
  {2012})}\BibitemShut {NoStop}%
\bibitem [{\citenamefont {K\'onya}\ \emph {et~al.}(2012)\citenamefont
  {K\'onya}, \citenamefont {Nagy}, \citenamefont {Szirmai},\ and\ \citenamefont
  {Domokos}}]{domokos_open_fs_2012}%
  \BibitemOpen
  \bibfield  {author} {\bibinfo {author} {\bibfnamefont {G.}~\bibnamefont
  {K\'onya}}, \bibinfo {author} {\bibfnamefont {D.}~\bibnamefont {Nagy}},
  \bibinfo {author} {\bibfnamefont {G.}~\bibnamefont {Szirmai}}, \ and\
  \bibinfo {author} {\bibfnamefont {P.}~\bibnamefont {Domokos}},\ }\Doi
  {10.1103/PhysRevA.86.013641} {\bibfield  {journal} {\bibinfo  {journal}
  {Phys. Rev. A},\ }\textbf {\bibinfo {volume} {86}},\ \bibinfo {pages}
  {013641} (\bibinfo {year} {2012})}\BibitemShut {NoStop}%
\bibitem [{\citenamefont {Nagy}\ \emph {et~al.}(2011)\citenamefont {Nagy},
  \citenamefont {Szirmai},\ and\ \citenamefont
  {Domokos}}]{domokos_open_ce_2011}%
  \BibitemOpen
  \bibfield  {author} {\bibinfo {author} {\bibfnamefont {D.}~\bibnamefont
  {Nagy}}, \bibinfo {author} {\bibfnamefont {G.}~\bibnamefont {Szirmai}}, \
  and\ \bibinfo {author} {\bibfnamefont {P.}~\bibnamefont {Domokos}},\ }\Doi
  {10.1103/PhysRevA.84.043637} {\bibfield  {journal} {\bibinfo  {journal}
  {Phys. Rev. A},\ }\textbf {\bibinfo {volume} {84}},\ \bibinfo {pages}
  {043637} (\bibinfo {year} {2011})}\BibitemShut {NoStop}%
\bibitem [{\citenamefont {Torre}\ \emph {et~al.}(2013)\citenamefont {Torre},
  \citenamefont {Diehl}, \citenamefont {Lukin}, \citenamefont {Sachdev},\ and\
  \citenamefont {Strack}}]{dallatorre_2013}%
  \BibitemOpen
  \bibfield  {author} {\bibinfo {author} {\bibfnamefont {E.~G.~D.}\
  \bibnamefont {Torre}}, \bibinfo {author} {\bibfnamefont {S.}~\bibnamefont
  {Diehl}}, \bibinfo {author} {\bibfnamefont {M.~D.}\ \bibnamefont {Lukin}},
  \bibinfo {author} {\bibfnamefont {S.}~\bibnamefont {Sachdev}}, \ and\
  \bibinfo {author} {\bibfnamefont {P.}~\bibnamefont {Strack}},\ }\Doi
  {10.1103/PhysRevA.87.023831} {\bibfield  {journal} {\bibinfo  {journal}
  {Phys. Rev. A},\ }\textbf {\bibinfo {volume} {87}},\ \bibinfo {pages}
  {023831} (\bibinfo {year} {2013})}\BibitemShut {NoStop}%
\bibitem [{\citenamefont {Tomka}\ \emph {et~al.}(2013)\citenamefont {Tomka},
  \citenamefont {Baeriswyl},\ and\ \citenamefont
  {Gritsev}}]{gritsev_driven_2013}%
  \BibitemOpen
  \bibfield  {author} {\bibinfo {author} {\bibfnamefont {M.}~\bibnamefont
  {Tomka}}, \bibinfo {author} {\bibfnamefont {D.}~\bibnamefont {Baeriswyl}}, \
  and\ \bibinfo {author} {\bibfnamefont {V.}~\bibnamefont {Gritsev}},\ }\Doi
  {10.1103/PhysRevA.88.053801} {\bibfield  {journal} {\bibinfo  {journal}
  {Phys. Rev. A},\ }\textbf {\bibinfo {volume} {88}},\ \bibinfo {pages}
  {053801} (\bibinfo {year} {2013})}\BibitemShut {NoStop}%
\bibitem [{\citenamefont {Genway}\ \emph {et~al.}(2014)\citenamefont {Genway},
  \citenamefont {Li}, \citenamefont {Ates}, \citenamefont {Lanyon},\ and\
  \citenamefont {Lesanovsky}}]{lesanovski_dicke_ions}%
  \BibitemOpen
  \bibfield  {author} {\bibinfo {author} {\bibfnamefont {S.}~\bibnamefont
  {Genway}}, \bibinfo {author} {\bibfnamefont {W.}~\bibnamefont {Li}}, \bibinfo
  {author} {\bibfnamefont {C.}~\bibnamefont {Ates}}, \bibinfo {author}
  {\bibfnamefont {B.~P.}\ \bibnamefont {Lanyon}}, \ and\ \bibinfo {author}
  {\bibfnamefont {I.}~\bibnamefont {Lesanovsky}},\ }\Doi
  {10.1103/PhysRevLett.112.023603} {\bibfield  {journal} {\bibinfo  {journal}
  {Phys. Rev. Lett.},\ }\textbf {\bibinfo {volume} {112}},\ \bibinfo {pages}
  {023603} (\bibinfo {year} {2014})}\BibitemShut {NoStop}%
\bibitem [{\citenamefont {Nagy}\ and\ \citenamefont
  {Domokos}(2015)}]{domokos_keldysh_2015}%
  \BibitemOpen
  \bibfield  {author} {\bibinfo {author} {\bibfnamefont {D.}~\bibnamefont
  {Nagy}}\ and\ \bibinfo {author} {\bibfnamefont {P.}~\bibnamefont {Domokos}},\
  }\Doi {10.1103/PhysRevLett.115.043601} {\bibfield  {journal} {\bibinfo
  {journal} {Phys. Rev. Lett.},\ }\textbf {\bibinfo {volume} {115}},\ \bibinfo
  {pages} {043601} (\bibinfo {year} {2015})}\BibitemShut {NoStop}%
\bibitem [{\citenamefont {Acevedo}\ \emph {et~al.}(2015)\citenamefont
  {Acevedo}, \citenamefont {Quiroga}, \citenamefont {Rodríguez},\ and\
  \citenamefont {Johnson}}]{acevedo_2015}%
  \BibitemOpen
  \bibfield  {author} {\bibinfo {author} {\bibfnamefont {O.~L.}\ \bibnamefont
  {Acevedo}}, \bibinfo {author} {\bibfnamefont {L.}~\bibnamefont {Quiroga}},
  \bibinfo {author} {\bibfnamefont {F.~J.}\ \bibnamefont {Rodríguez}}, \ and\
  \bibinfo {author} {\bibfnamefont {N.~F.}\ \bibnamefont {Johnson}},\ }\href
  {http://stacks.iop.org/1367-2630/17/i=9/a=093005} {\bibfield  {journal}
  {\bibinfo  {journal} {New Journal of Physics},\ }\textbf {\bibinfo {volume}
  {17}},\ \bibinfo {pages} {093005} (\bibinfo {year} {2015})}\BibitemShut
  {NoStop}%
\bibitem [{\citenamefont {Dicke}(1954)}]{dicke_54}%
  \BibitemOpen
  \bibfield  {author} {\bibinfo {author} {\bibfnamefont {R.~H.}\ \bibnamefont
  {Dicke}},\ }\Doi {10.1103/PhysRev.93.99} {\bibfield  {journal} {\bibinfo
  {journal} {Phys. Rev.},\ }\textbf {\bibinfo {volume} {93}},\ \bibinfo {pages}
  {99} (\bibinfo {year} {1954})}\BibitemShut {NoStop}%
\bibitem [{\citenamefont {Hepp}\ and\ \citenamefont {Lieb}(1973)}]{lieb_1973}%
  \BibitemOpen
  \bibfield  {author} {\bibinfo {author} {\bibfnamefont {K.}~\bibnamefont
  {Hepp}}\ and\ \bibinfo {author} {\bibfnamefont {E.~H.}\ \bibnamefont
  {Lieb}},\ }\Doi {http://dx.doi.org/10.1016/0003-4916(73)90039-0} {\bibfield
  {journal} {\bibinfo  {journal} {Annals of Physics},\ }\textbf {\bibinfo
  {volume} {76}},\ \bibinfo {pages} {360 } (\bibinfo {year} {1973})},\ ISSN
  \bibinfo {issn} {0003-4916}\BibitemShut {NoStop}%
\bibitem [{\citenamefont {Wang}\ and\ \citenamefont {Hioe}(1973)}]{hioe_1973}%
  \BibitemOpen
  \bibfield  {author} {\bibinfo {author} {\bibfnamefont {Y.~K.}\ \bibnamefont
  {Wang}}\ and\ \bibinfo {author} {\bibfnamefont {F.~T.}\ \bibnamefont
  {Hioe}},\ }\Doi {10.1103/PhysRevA.7.831} {\bibfield  {journal} {\bibinfo
  {journal} {Phys. Rev. A},\ }\textbf {\bibinfo {volume} {7}},\ \bibinfo
  {pages} {831} (\bibinfo {year} {1973})}\BibitemShut {NoStop}%
\bibitem [{\citenamefont {Lambert}\ \emph {et~al.}(2004)\citenamefont
  {Lambert}, \citenamefont {Emary},\ and\ \citenamefont
  {Brandes}}]{emary_2004}%
  \BibitemOpen
  \bibfield  {author} {\bibinfo {author} {\bibfnamefont {N.}~\bibnamefont
  {Lambert}}, \bibinfo {author} {\bibfnamefont {C.}~\bibnamefont {Emary}}, \
  and\ \bibinfo {author} {\bibfnamefont {T.}~\bibnamefont {Brandes}},\ }\Doi
  {10.1103/PhysRevLett.92.073602} {\bibfield  {journal} {\bibinfo  {journal}
  {Phys. Rev. Lett.},\ }\textbf {\bibinfo {volume} {92}},\ \bibinfo {pages}
  {073602} (\bibinfo {year} {2004})}\BibitemShut {NoStop}%
\bibitem [{\citenamefont {Vidal}\ and\ \citenamefont
  {Dusuel}(2006)}]{vidal_2006}%
  \BibitemOpen
  \bibfield  {author} {\bibinfo {author} {\bibfnamefont {J.}~\bibnamefont
  {Vidal}}\ and\ \bibinfo {author} {\bibfnamefont {S.}~\bibnamefont {Dusuel}},\
  }\href@noop {} {\bibfield  {journal} {\bibinfo  {journal} {EPL (Europhysics
  Letters)},\ }\textbf {\bibinfo {volume} {74}},\ \bibinfo {pages} {817}
  (\bibinfo {year} {2006})}\BibitemShut {NoStop}%
\bibitem [{\citenamefont {Liu}\ \emph {et~al.}(2009)\citenamefont {Liu},
  \citenamefont {Zhang}, \citenamefont {Chen},\ and\ \citenamefont
  {Wang}}]{liu_finite_2009}%
  \BibitemOpen
  \bibfield  {author} {\bibinfo {author} {\bibfnamefont {T.}~\bibnamefont
  {Liu}}, \bibinfo {author} {\bibfnamefont {Y.-Y.}\ \bibnamefont {Zhang}},
  \bibinfo {author} {\bibfnamefont {Q.-H.}\ \bibnamefont {Chen}}, \ and\
  \bibinfo {author} {\bibfnamefont {K.-L.}\ \bibnamefont {Wang}},\ }\Doi
  {10.1103/PhysRevA.80.023810} {\bibfield  {journal} {\bibinfo  {journal}
  {Phys. Rev. A},\ }\textbf {\bibinfo {volume} {80}},\ \bibinfo {pages}
  {023810} (\bibinfo {year} {2009})}\BibitemShut {NoStop}%
\bibitem [{\citenamefont {Larson}\ and\ \citenamefont
  {Lewenstein}(2009)}]{larson_lew_2009}%
  \BibitemOpen
  \bibfield  {author} {\bibinfo {author} {\bibfnamefont {J.}~\bibnamefont
  {Larson}}\ and\ \bibinfo {author} {\bibfnamefont {M.}~\bibnamefont
  {Lewenstein}},\ }\href@noop {} {\bibfield  {journal} {\bibinfo  {journal}
  {New Journal of Physics},\ }\textbf {\bibinfo {volume} {11}},\ \bibinfo
  {pages} {063027} (\bibinfo {year} {2009})}\BibitemShut {NoStop}%
\bibitem [{\citenamefont {Nagy}\ \emph {et~al.}(2010)\citenamefont {Nagy},
  \citenamefont {K\'onya}, \citenamefont {Szirmai},\ and\ \citenamefont
  {Domokos}}]{nagy_2010}%
  \BibitemOpen
  \bibfield  {author} {\bibinfo {author} {\bibfnamefont {D.}~\bibnamefont
  {Nagy}}, \bibinfo {author} {\bibfnamefont {G.}~\bibnamefont {K\'onya}},
  \bibinfo {author} {\bibfnamefont {G.}~\bibnamefont {Szirmai}}, \ and\
  \bibinfo {author} {\bibfnamefont {P.}~\bibnamefont {Domokos}},\ }\Doi
  {10.1103/PhysRevLett.104.130401} {\bibfield  {journal} {\bibinfo  {journal}
  {Phys. Rev. Lett.},\ }\textbf {\bibinfo {volume} {104}},\ \bibinfo {pages}
  {130401} (\bibinfo {year} {2010})}\BibitemShut {NoStop}%
\bibitem [{\citenamefont {Mumford}\ \emph {et~al.}(2014)\citenamefont
  {Mumford}, \citenamefont {Larson},\ and\ \citenamefont
  {O'Dell}}]{larson_imp_2014}%
  \BibitemOpen
  \bibfield  {author} {\bibinfo {author} {\bibfnamefont {J.}~\bibnamefont
  {Mumford}}, \bibinfo {author} {\bibfnamefont {J.}~\bibnamefont {Larson}}, \
  and\ \bibinfo {author} {\bibfnamefont {D.~H.~J.}\ \bibnamefont {O'Dell}},\
  }\Doi {10.1103/PhysRevA.89.023620} {\bibfield  {journal} {\bibinfo  {journal}
  {Phys. Rev. A},\ }\textbf {\bibinfo {volume} {89}},\ \bibinfo {pages}
  {023620} (\bibinfo {year} {2014})}\BibitemShut {NoStop}%
\bibitem [{\citenamefont {Baumann}\ \emph {et~al.}(2010)\citenamefont
  {Baumann}, \citenamefont {Guerlin}, \citenamefont {Brennecke},\ and\
  \citenamefont {Esslinger}}]{eth_2010}%
  \BibitemOpen
  \bibfield  {author} {\bibinfo {author} {\bibfnamefont {K.}~\bibnamefont
  {Baumann}}, \bibinfo {author} {\bibfnamefont {C.}~\bibnamefont {Guerlin}},
  \bibinfo {author} {\bibfnamefont {F.}~\bibnamefont {Brennecke}}, \ and\
  \bibinfo {author} {\bibfnamefont {T.}~\bibnamefont {Esslinger}},\ }\href@noop
  {} {\bibfield  {journal} {\bibinfo  {journal} {Nature},\ }\textbf {\bibinfo
  {volume} {464}},\ \bibinfo {pages} {1301} (\bibinfo {year}
  {2010})}\BibitemShut {NoStop}%
\bibitem [{\citenamefont {Baumann}\ \emph {et~al.}(2011)\citenamefont
  {Baumann}, \citenamefont {Mottl}, \citenamefont {Brennecke},\ and\
  \citenamefont {Esslinger}}]{eth_2011}%
  \BibitemOpen
  \bibfield  {author} {\bibinfo {author} {\bibfnamefont {K.}~\bibnamefont
  {Baumann}}, \bibinfo {author} {\bibfnamefont {R.}~\bibnamefont {Mottl}},
  \bibinfo {author} {\bibfnamefont {F.}~\bibnamefont {Brennecke}}, \ and\
  \bibinfo {author} {\bibfnamefont {T.}~\bibnamefont {Esslinger}},\ }\Doi
  {10.1103/PhysRevLett.107.140402} {\bibfield  {journal} {\bibinfo  {journal}
  {Phys. Rev. Lett.},\ }\textbf {\bibinfo {volume} {107}},\ \bibinfo {pages}
  {140402} (\bibinfo {year} {2011})}\BibitemShut {NoStop}%
\bibitem [{\citenamefont {Brennecke}\ \emph {et~al.}(2013)\citenamefont
  {Brennecke}, \citenamefont {Mottl}, \citenamefont {Baumann}, \citenamefont
  {Landig}, \citenamefont {Donner},\ and\ \citenamefont
  {Esslinger}}]{eth_2013}%
  \BibitemOpen
  \bibfield  {author} {\bibinfo {author} {\bibfnamefont {F.}~\bibnamefont
  {Brennecke}}, \bibinfo {author} {\bibfnamefont {R.}~\bibnamefont {Mottl}},
  \bibinfo {author} {\bibfnamefont {K.}~\bibnamefont {Baumann}}, \bibinfo
  {author} {\bibfnamefont {R.}~\bibnamefont {Landig}}, \bibinfo {author}
  {\bibfnamefont {T.}~\bibnamefont {Donner}}, \ and\ \bibinfo {author}
  {\bibfnamefont {T.}~\bibnamefont {Esslinger}},\ }\Doi
  {10.1073/pnas.1306993110} {\bibfield  {journal} {\bibinfo  {journal}
  {Proceedings of the National Academy of Sciences},\ }\textbf {\bibinfo
  {volume} {110}},\ \bibinfo {pages} {11763} (\bibinfo {year}
  {2013})}\BibitemShut {NoStop}%
\bibitem [{\citenamefont {Baden}\ \emph {et~al.}(2014)\citenamefont {Baden},
  \citenamefont {Arnold}, \citenamefont {Grimsmo}, \citenamefont {Parkins},\
  and\ \citenamefont {Barrett}}]{barrett_2014}%
  \BibitemOpen
  \bibfield  {author} {\bibinfo {author} {\bibfnamefont {M.~P.}\ \bibnamefont
  {Baden}}, \bibinfo {author} {\bibfnamefont {K.~J.}\ \bibnamefont {Arnold}},
  \bibinfo {author} {\bibfnamefont {A.~L.}\ \bibnamefont {Grimsmo}}, \bibinfo
  {author} {\bibfnamefont {S.}~\bibnamefont {Parkins}}, \ and\ \bibinfo
  {author} {\bibfnamefont {M.~D.}\ \bibnamefont {Barrett}},\ }\Doi
  {10.1103/PhysRevLett.113.020408} {\bibfield  {journal} {\bibinfo  {journal}
  {Phys. Rev. Lett.},\ }\textbf {\bibinfo {volume} {113}},\ \bibinfo {pages}
  {020408} (\bibinfo {year} {2014})}\BibitemShut {NoStop}%
\bibitem [{\citenamefont {Ke\ss{}ler}\ \emph {et~al.}(2014)\citenamefont
  {Ke\ss{}ler}, \citenamefont {Klinder}, \citenamefont {Wolke},\ and\
  \citenamefont {Hemmerich}}]{hemmerich_2014}%
  \BibitemOpen
  \bibfield  {author} {\bibinfo {author} {\bibfnamefont {H.}~\bibnamefont
  {Ke\ss{}ler}}, \bibinfo {author} {\bibfnamefont {J.}~\bibnamefont {Klinder}},
  \bibinfo {author} {\bibfnamefont {M.}~\bibnamefont {Wolke}}, \ and\ \bibinfo
  {author} {\bibfnamefont {A.}~\bibnamefont {Hemmerich}},\ }\Doi
  {10.1103/PhysRevLett.113.070404} {\bibfield  {journal} {\bibinfo  {journal}
  {Phys. Rev. Lett.},\ }\textbf {\bibinfo {volume} {113}},\ \bibinfo {pages}
  {070404} (\bibinfo {year} {2014})}\BibitemShut {NoStop}%
\bibitem [{\citenamefont {Klinder}\ \emph {et~al.}(2015)\citenamefont
  {Klinder}, \citenamefont {Keßler}, \citenamefont {Wolke}, \citenamefont
  {Mathey},\ and\ \citenamefont {Hemmerich}}]{hemmerich_2015}%
  \BibitemOpen
  \bibfield  {author} {\bibinfo {author} {\bibfnamefont {J.}~\bibnamefont
  {Klinder}}, \bibinfo {author} {\bibfnamefont {H.}~\bibnamefont {Keßler}},
  \bibinfo {author} {\bibfnamefont {M.}~\bibnamefont {Wolke}}, \bibinfo
  {author} {\bibfnamefont {L.}~\bibnamefont {Mathey}}, \ and\ \bibinfo {author}
  {\bibfnamefont {A.}~\bibnamefont {Hemmerich}},\ }\Doi
  {10.1073/pnas.1417132112} {\bibfield  {journal} {\bibinfo  {journal}
  {Proceedings of the National Academy of Sciences},\ }\textbf {\bibinfo
  {volume} {112}},\ \bibinfo {pages} {3290} (\bibinfo {year}
  {2015})}\BibitemShut {NoStop}%
\bibitem [{\citenamefont {Fagotti}\ and\ \citenamefont
  {Essler}(2013)}]{essler_2013}%
  \BibitemOpen
  \bibfield  {author} {\bibinfo {author} {\bibfnamefont {M.}~\bibnamefont
  {Fagotti}}\ and\ \bibinfo {author} {\bibfnamefont {F.~H.~L.}\ \bibnamefont
  {Essler}},\ }\Doi {10.1103/PhysRevB.87.245107} {\bibfield  {journal}
  {\bibinfo  {journal} {Phys. Rev. B},\ }\textbf {\bibinfo {volume} {87}},\
  \bibinfo {pages} {245107} (\bibinfo {year} {2013})}\BibitemShut {NoStop}%
\bibitem [{\citenamefont {Emary}\ and\ \citenamefont
  {Brandes}(2003)}]{emary_2003}%
  \BibitemOpen
  \bibfield  {author} {\bibinfo {author} {\bibfnamefont {C.}~\bibnamefont
  {Emary}}\ and\ \bibinfo {author} {\bibfnamefont {T.}~\bibnamefont
  {Brandes}},\ }\Doi {10.1103/PhysRevE.67.066203} {\bibfield  {journal}
  {\bibinfo  {journal} {Phys. Rev. E},\ }\textbf {\bibinfo {volume} {67}},\
  \bibinfo {pages} {066203} (\bibinfo {year} {2003})}\BibitemShut {NoStop}%
\bibitem [{\citenamefont {Bakemeier}\ \emph {et~al.}(2013)\citenamefont
  {Bakemeier}, \citenamefont {Alvermann},\ and\ \citenamefont
  {Fehske}}]{fehske_2013}%
  \BibitemOpen
  \bibfield  {author} {\bibinfo {author} {\bibfnamefont {L.}~\bibnamefont
  {Bakemeier}}, \bibinfo {author} {\bibfnamefont {A.}~\bibnamefont
  {Alvermann}}, \ and\ \bibinfo {author} {\bibfnamefont {H.}~\bibnamefont
  {Fehske}},\ }\Doi {10.1103/PhysRevA.88.043835} {\bibfield  {journal}
  {\bibinfo  {journal} {Phys. Rev. A},\ }\textbf {\bibinfo {volume} {88}},\
  \bibinfo {pages} {043835} (\bibinfo {year} {2013})}\BibitemShut {NoStop}%
\bibitem [{\citenamefont {Altland}\ and\ \citenamefont
  {Haake}(2012)}]{haake_2012}%
  \BibitemOpen
  \bibfield  {author} {\bibinfo {author} {\bibfnamefont {A.}~\bibnamefont
  {Altland}}\ and\ \bibinfo {author} {\bibfnamefont {F.}~\bibnamefont
  {Haake}},\ }\Doi {10.1103/PhysRevLett.108.073601} {\bibfield  {journal}
  {\bibinfo  {journal} {Phys. Rev. Lett.},\ }\textbf {\bibinfo {volume}
  {108}},\ \bibinfo {pages} {073601} (\bibinfo {year} {2012})}\BibitemShut
  {NoStop}%
\bibitem [{\citenamefont {Buchhold}\ \emph {et~al.}(2013)\citenamefont
  {Buchhold}, \citenamefont {Strack}, \citenamefont {Sachdev},\ and\
  \citenamefont {Diehl}}]{buchhold_dis_2013}%
  \BibitemOpen
  \bibfield  {author} {\bibinfo {author} {\bibfnamefont {M.}~\bibnamefont
  {Buchhold}}, \bibinfo {author} {\bibfnamefont {P.}~\bibnamefont {Strack}},
  \bibinfo {author} {\bibfnamefont {S.}~\bibnamefont {Sachdev}}, \ and\
  \bibinfo {author} {\bibfnamefont {S.}~\bibnamefont {Diehl}},\ }\Doi
  {10.1103/PhysRevA.87.063622} {\bibfield  {journal} {\bibinfo  {journal}
  {Phys. Rev. A},\ }\textbf {\bibinfo {volume} {87}},\ \bibinfo {pages}
  {063622} (\bibinfo {year} {2013})}\BibitemShut {NoStop}%
\bibitem [{\citenamefont {Kulkarni}\ \emph {et~al.}(2013)\citenamefont
  {Kulkarni}, \citenamefont {\"Oztop},\ and\ \citenamefont
  {T\"ureci}}]{tureci_nearss}%
  \BibitemOpen
  \bibfield  {author} {\bibinfo {author} {\bibfnamefont {M.}~\bibnamefont
  {Kulkarni}}, \bibinfo {author} {\bibfnamefont {B.}~\bibnamefont {\"Oztop}}, \
  and\ \bibinfo {author} {\bibfnamefont {H.~E.}\ \bibnamefont {T\"ureci}},\
  }\Doi {10.1103/PhysRevLett.111.220408} {\bibfield  {journal} {\bibinfo
  {journal} {Phys. Rev. Lett.},\ }\textbf {\bibinfo {volume} {111}},\ \bibinfo
  {pages} {220408} (\bibinfo {year} {2013})}\BibitemShut {NoStop}%
\bibitem [{\citenamefont {K\'onya}\ \emph {et~al.}(2014)\citenamefont
  {K\'onya}, \citenamefont {Szirmai},\ and\ \citenamefont
  {Domokos}}]{domokos_damping}%
  \BibitemOpen
  \bibfield  {author} {\bibinfo {author} {\bibfnamefont {G.}~\bibnamefont
  {K\'onya}}, \bibinfo {author} {\bibfnamefont {G.}~\bibnamefont {Szirmai}}, \
  and\ \bibinfo {author} {\bibfnamefont {P.}~\bibnamefont {Domokos}},\ }\Doi
  {10.1103/PhysRevA.90.013623} {\bibfield  {journal} {\bibinfo  {journal}
  {Phys. Rev. A},\ }\textbf {\bibinfo {volume} {90}},\ \bibinfo {pages}
  {013623} (\bibinfo {year} {2014})}\BibitemShut {NoStop}%
\bibitem [{\citenamefont {Kamenev}(2011)}]{kamenev_book}%
  \BibitemOpen
  \bibfield  {author} {\bibinfo {author} {\bibfnamefont {A.}~\bibnamefont
  {Kamenev}},\ }\href {http://books.google.de/books?id=CwlrUepnla4C} {\emph
  {\bibinfo {title} {Field Theory of Non-Equilibrium Systems}}}\ (\bibinfo
  {publisher} {Cambridge University Press},\ \bibinfo {year} {2011})\ ISBN
  \bibinfo {isbn} {9781139500296}\BibitemShut {NoStop}%
\bibitem [{\citenamefont {Sieberer}\ \emph {et~al.}(2015)\citenamefont
  {Sieberer}, \citenamefont {Buchhold},\ and\ \citenamefont
  {Diehl}}]{sieberer2015keldysh}%
  \BibitemOpen
  \bibfield  {author} {\bibinfo {author} {\bibfnamefont {L.}~\bibnamefont
  {Sieberer}}, \bibinfo {author} {\bibfnamefont {M.}~\bibnamefont {Buchhold}},
  \ and\ \bibinfo {author} {\bibfnamefont {S.}~\bibnamefont {Diehl}},\
  }\href@noop {} {\bibfield  {journal} {\bibinfo  {journal} {arXiv preprint
  arXiv:1512.00637}} (\bibinfo {year} {2015})}\BibitemShut {NoStop}%
\bibitem [{Note1()}]{Note1}%
  \BibitemOpen
  \bibinfo {note} {See Supplemental Material}\BibitemShut {NoStop}%
\bibitem [{\citenamefont {Marcuzzi}\ \emph {et~al.}(2013)\citenamefont
  {Marcuzzi}, \citenamefont {Marino}, \citenamefont {Gambassi},\ and\
  \citenamefont {Silva}}]{marino_2013}%
  \BibitemOpen
  \bibfield  {author} {\bibinfo {author} {\bibfnamefont {M.}~\bibnamefont
  {Marcuzzi}}, \bibinfo {author} {\bibfnamefont {J.}~\bibnamefont {Marino}},
  \bibinfo {author} {\bibfnamefont {A.}~\bibnamefont {Gambassi}}, \ and\
  \bibinfo {author} {\bibfnamefont {A.}~\bibnamefont {Silva}},\ }\Doi
  {10.1103/PhysRevLett.111.197203} {\bibfield  {journal} {\bibinfo  {journal}
  {Phys. Rev. Lett.},\ }\textbf {\bibinfo {volume} {111}},\ \bibinfo {pages}
  {197203} (\bibinfo {year} {2013})}\BibitemShut {NoStop}%
\bibitem [{Note2()}]{Note2}%
  \BibitemOpen
  \bibinfo {note} {Note that the single-mode approximation, implicit in the
  results presented here, breaks down for small coupling constants, where
  $\omega _\protect \text {qp}\not =0$ and the two most relevant modes have
  degenerate lifetimes.}\BibitemShut {Stop}%
\bibitem [{Note3()}]{Note3}%
  \BibitemOpen
  \bibinfo {note} {The scaling laws we found for $\kappa _{\protect \rm
  kin},\kappa _{\protect \rm qp},\lambda _{\protect \rm kin}$, and $g_c(N)$ can
  be also derived from the scale-invariance of the GFs, which holds in the
  strong-coupling region. See J. Lang and F. Piazza, in preparation
  (2016)}\BibitemShut {NoStop}%
\bibitem [{\citenamefont {Foini}\ \emph {et~al.}(2011)\citenamefont {Foini},
  \citenamefont {Cugliandolo},\ and\ \citenamefont {Gambassi}}]{gambassi_2011}%
  \BibitemOpen
  \bibfield  {author} {\bibinfo {author} {\bibfnamefont {L.}~\bibnamefont
  {Foini}}, \bibinfo {author} {\bibfnamefont {L.~F.}\ \bibnamefont
  {Cugliandolo}}, \ and\ \bibinfo {author} {\bibfnamefont {A.}~\bibnamefont
  {Gambassi}},\ }\Doi {10.1103/PhysRevB.84.212404} {\bibfield  {journal}
  {\bibinfo  {journal} {Phys. Rev. B},\ }\textbf {\bibinfo {volume} {84}},\
  \bibinfo {pages} {212404} (\bibinfo {year} {2011})}\BibitemShut {NoStop}%
\bibitem [{\citenamefont {Sciolla}\ and\ \citenamefont
  {Biroli}(2013)}]{sciolla_2013}%
  \BibitemOpen
  \bibfield  {author} {\bibinfo {author} {\bibfnamefont {B.}~\bibnamefont
  {Sciolla}}\ and\ \bibinfo {author} {\bibfnamefont {G.}~\bibnamefont
  {Biroli}},\ }\Doi {10.1103/PhysRevB.88.201110} {\bibfield  {journal}
  {\bibinfo  {journal} {Phys. Rev. B},\ }\textbf {\bibinfo {volume} {88}},\
  \bibinfo {pages} {201110} (\bibinfo {year} {2013})}\BibitemShut {NoStop}%
\bibitem [{\citenamefont {Chiocchetta}\ \emph {et~al.}(2015)\citenamefont
  {Chiocchetta}, \citenamefont {Tavora}, \citenamefont {Gambassi},\ and\
  \citenamefont {Mitra}}]{ciocchetta_2015}%
  \BibitemOpen
  \bibfield  {author} {\bibinfo {author} {\bibfnamefont {A.}~\bibnamefont
  {Chiocchetta}}, \bibinfo {author} {\bibfnamefont {M.}~\bibnamefont {Tavora}},
  \bibinfo {author} {\bibfnamefont {A.}~\bibnamefont {Gambassi}}, \ and\
  \bibinfo {author} {\bibfnamefont {A.}~\bibnamefont {Mitra}},\ }\Doi
  {10.1103/PhysRevB.91.220302} {\bibfield  {journal} {\bibinfo  {journal}
  {Phys. Rev. B},\ }\textbf {\bibinfo {volume} {91}},\ \bibinfo {pages}
  {220302} (\bibinfo {year} {2015})}\BibitemShut {NoStop}%
\bibitem [{Note4()}]{Note4}%
  \BibitemOpen
  \bibinfo {note} {J. Lang and F. Piazza, in preparation (2016)}\BibitemShut
  {NoStop}%
\end{thebibliography}%


\end{document}